\documentclass[11pt]{article}
\makeatletter
\renewcommand\section{\@startsection{section}{1}{\z@}%
                                    {-3.25ex\@plus -1ex \@minus -.2ex}%
                                    {1.5ex \@plus .2ex}%
                                    {\reset@font\large\bfseries}}
\renewcommand\subsection{\@startsection{subsection}{2}{\z@}%
                                    {3.25ex \@plus 1ex \@minus.2ex}%
                                    {-1em}%
                                    {\reset@font\normalsize\bfseries}}
\@addtoreset{equation}{section}
\makeatother
\renewcommand{\theequation}{\thesection.\arabic{equation}}
\def\be{\begin{equation}\label}            \def\ba{\begin{eqnarray}}
\def\ee{\end{equation}}                    \def\ea{\end{eqnarray}}
\newcommand{\ar}{\begin{array}}            \newcommand{\er}{\end{array}}
\def\nn{\nonumber}                         

\def\={\,=}           \def\D{\Delta }         
          
\begin{document}
\hyphenation{ge-ne-ra-tor ge-ne-ra-tors  mo-du-lar}
\setcounter{page}{0}
\thispagestyle{empty}
September~~1998
\hfill{\begin{tabular}{l}  
       \sc hep-th/9809001 \\
       \sc Sfb-288 preprint 343\\
       \end{tabular} }
\vspace*{1.7cm}

\begin{center}
{\Large\bf Factorized Combinations of Virasoro Characters } \\ [1.1cm]
{\sc Andrei G. Bytsko}~${}^{1,2}$,\ \ \ {\sc Andreas Fring}~${}^{1}$
  \\ [4mm]
${}^{1}$
Institut f\"ur Theoretische Physik, Freie Universit\"at Berlin \\
Arnimallee 14, 14195 Berlin, Germany \\ [2.5mm]
${}^{2}$
Steklov Mathematical Institute, Fontanka 27,\\
 St.Petersburg~~191011, Russia \\ [30mm]
{\bf Abstract} \\ [9mm]

\parbox{13cm}{ We investigate linear combinations of characters for 
minimal Virasoro models which are representable as a product
of several basic blocks. 
Our analysis is based on consideration of asymptotic 
behaviour of the characters in the quasi-classical limit.
In particular, we introduce a notion of the secondary 
effective central charge.
We find all possible cases for which factorization occurs 
on the base of the Gau{\ss}-Jacobi or the Watson identities. 
Exploiting these results, we establish various types of
identities between different characters. In particular,
we present several identities generalizing the Rogers-Ramanujan
identities.
Applications to quasi-particle representations, modular
invariant partition functions, super-conformal theories
and conformal models with boundaries are briefly discussed.
}  \end{center}

\vfill{ \hspace*{-9mm}
 \begin{tabular}{l}
\rule{6 cm}{0.05 mm} \\
 e-mail:\   bytsko@pdmi.ras.ru ,\ fring@physik.fu-berlin.de
 \end{tabular} }
\setcounter{section}{0}

\newpage

\section*{Introduction} \ 

\noindent
It is a well known fact that the characters of
irreducible representations of the Virasoro algebra for the 
${\cal M}(3,4)$ minimal model possess the peculiar property to be
representable as infinite products
\ba 
 &   \label{e1}
 \chi_{1,2}^{3,4}(q)  \= q^{\frac{1}{24}} 
 \prod\limits_{n=0}^\infty (1+q^{n+1}) \= 
  q^{\frac{1}{24}} \prod\limits_{n=0}^\infty 
 \Bigl(\frac{1}{1-q^{2n+1}} \Bigr) \, , & \\
 &  \label{e2}
 \chi_{1,1}^{3,4}(q) \pm \chi_{1,3}^{3,4}(q)  \= q^{-\frac{1}{48}} 
 \prod\limits_{n=0}^\infty (1 \pm q^{n+1/2}) \, . &
\ea
As was observed in \cite{Ro}, some characters and linear 
combinations of characters for the ${\cal M}(4,5)$ minimal model
admit similar forms.

The question towards a generalization and classification of these
identities arises naturally. 
Surprisingly, it turned out \cite{chr} that the only factorizable 
single characters for minimal models are of type 
$\chi_{n,m}^{2n,t}(q)$ and $\chi_{n,m}^{3n,t}(q)$.
In \cite{cap,chr,KRV,ES,BF1} it was discussed that the factorization 
of characters in these series is based on the Gau{\ss}-Jacobi and 
Watson identities. 

On the other hand, a factorization of linear combinations of
Virasoro characters has not been studied so far.
In the present paper we show that factorization of
combinations $\chi_{n,m}^{s,t}(q) \pm \chi_{n,t-m}^{s,t}(q)$
occurring due to the Gau{\ss}-Jacobi and Watson identities
is possible (up to the symmetries of the characters) 
only for $s=3n,4n,6n$. Moreover, we will prove that
there are no other factorizable differences of this type
which admit the inverse product form similar to the r.h.s. of
(\ref{e1}).

We present a systematic analysis based on considerations of the 
asymptotic behaviour of (combinations of) characters in the
so-called quasi-classical limit, $q \rightarrow 1^-$. 
We will demonstrate that  for linear combinations of the above 
mentioned type we need, besides the effective 
central charge $c_{\rm eff}$, the notion of the ``secondary'' 
effective central charge $\tilde{c}$. 

The advantage to have the characters (or combinations) in the form 
of infinite products rather than infinite sums is many-fold.
First of all the problem of finding the dimension of a particular 
level in the Verma module of the irreducible representation 
has been reduced to a simple problem of partitions.
As a consequence one may state the possible monomials of Virasoro
generators at a specific level.  Also the associated quasi-particle
states may be constructed from this form without any effort,
whereas it is virtually impossible to find them from the infinite 
sum representation. The quasi-particle form is also related
to a classification of Rogers-Ramanujan type of identities \cite{RSR}.
In the present paper this subject is discussed rather briefly in 
subsection 3.4 and appendix E. However, this point is followed up 
further in \cite{AB}, where the obtained factorized forms of
characters were exploited in the derivation of Rogers-Ramanujan type
identities.
In addition, the factorized characters (or combinations) allow 
to derive various new identities between different 
combinations of characters far easier than employing the
infinite sum representation. Some of these identities relate
different sectors of the same models, whereas others relate
different models altogether. Factorized combinations of characters
appear naturally in the context
of coset models, super-conformal extensions of the Virasoro 
algebra and boundary conformal field theories. 
They may even shed some light on massive models, since it was 
conjectured in \cite{chr} that they allow to identify the 
space of form factors of descendant operators.
\pagebreak

\section{Preliminaries} \ 

\noindent
We use the notation $\langle n,m\rangle=1$ if 
$n$ and $m$ are co-prime numbers and we employ also
the standard abbreviation for Euler's function 
$(q)_{m}=\prod\limits_{k=1}^{m}(1-q^{k})$ with $(q)_{0}=1$. 

\subsection{Characters of Minimal Models} \

\noindent
The Virasoro algebra is generated by operator valued
Fourier coefficients of the energy-momentum tensor  
$T(z)=\sum_n z^{-n-2} L_{n}$ and a central charge $c$.
For an irreducible highest weight representation $V_{c,h}$ of the
Virasoro algebra with central charge $c$ and weight $h$ one
defines the character
\be{chi}
 \chi(q) \= {\rm tr}_{V_{c,h}} q^{L_0 - \frac{c}{24} } \= 
 q^{h-\frac{c}{24} } \sum_{n=0}^{\infty} \mu_n \, q^n \, ,
\ee
with $\mu_n$ being the multiplicity of the level $n$. The corresponding 
states at a particular level $k$ are spanned by the 
vectors
\be{states}
L_{-k_1} \dots  L_{-k_n} | h \rangle,  \qquad k_1 \leq 
k_2 \leq \ldots  \leq k_n \,, \qquad k=\sum_{i=1}^n k_i   \, . 
\ee

Minimal models are the distinguished conformal theories 
in which the set of highest weights is finite \cite{BPZ}. These models are
labeled by two integer numbers $s$ and $t$ such that
\be{st}
  s, t \geq 2 \ \ \ \ \hbox{and} \ \ \ \ 
 \langle s,t \rangle = 1 \, .
\ee
The minimal models for which $|s-t|=1$ are unitary \cite{FQS,GKO}.
The minimal model ${\cal M}(s,t)$ has the central charge
\be{c} 
 c(s,t) \= 1 - \frac{6(s-t)^{2}}{s\;t} \, .
 \ee
The corresponding irreducible highest weight  representations 
of the Virasoro algebra are representations with the weights
\be{h} 
 h^{s,t}_{n,m}=\frac{(nt-ms)^{2}-(s-t)^{2}}{4\;s\;t} \ ,
\ee
where the labels run through the following set of integers:
\be{nm}
 1\leq n\leq s-1 \ , \ \ \ \ \ 1\leq m\leq t-1 \ . 
\ee
The corresponding character is given by \cite{FF,Ro}
\be{char}
 \chi_{n,m}^{s,t}(q) \= 
 \frac{ q^{h_{n,m}^{s,t} - \frac{c(s,t)}{24}} }{ (q)_\infty }
 \sum_{k=-\infty }^{\infty }q^{st k^2}
 \left(q^{k(nt-ms)}-q^{k(nt+ms)+nm}\right) \=
 \frac{q^{h_{n,m}^{s,t} - \frac{c(s,t)}{24}}}{(q)_{\infty }}
 \hat{\chi}_{n,m}^{s,t}(q) \, ,
\ee
(the second equality defines $\hat{\chi}(q)$,
which we refer to as ``incomplete character'').
The characters possess the following symmetries:
\be{hsym} 
 \chi^{s,t}_{n,m}(q) \= \chi^{t,s}_{m,n}(q) \= 
 \chi^{s,t}_{s-n,t-m}(q)  \= \chi^{t,s}_{t-m,s-n}(q) \, .
\ee
It follows from (\ref{nm}) and these symmetries that the minimal
model ${\cal M}(s,t)$ has ${\cal D}=(s-1)(t-1)/2$ different sectors
(inequivalent irreducible representations). In addition, (\ref{char})
allows to relate some characters of different models
\be{alp}
   \chi^{\alpha s,t}_{\alpha n,m}(q) \= 
   \chi^{s, \alpha t}_{n, \alpha m}(q) \, ,
\ee
where $\alpha$ is a positive number such that
$\langle s, \alpha t \rangle = \langle t, \alpha s \rangle =1$. 
For instance, $\chi^{6,5}_{2,m}(q) = \chi^{3,10}_{1,2m}(q)$.

\subsection{Quantum Dilogarithm} \

\noindent
In our analysis of factorized characters we will be exploiting the
properties of the quantum dilogarithm, whose defining relations are 
\be{quprod}
 \ln_{q}(\theta ) := \prod\limits_{k=0}^\infty
 (1-e^{2\pi i\theta}q^{k}) \=
 \exp \sum\limits_{k=1}^\infty 
 \frac{1}{k}\frac{e^{2\pi i\theta k}}{q^{k}-1} \, .  
\end{equation}
Taking $q=e^{2\pi i\tau },$ we assume that ${\rm Im}(\tau )>0$ and 
${\rm Im}(\theta)>0$ in order to guarantee the convergence of 
(\ref{quprod}). We see from (\ref{quprod}) that $\ln_{q}(\theta)$ 
is a pseudo-double-periodic function 
\be{qf1}
 \ln_{q}(\theta +1) \= 
 \ln _{q}(\theta ) \qquad \hbox{and} \qquad \ln_{q}
 \left(\theta +\tau \right) \=
 \frac{1}{1-e^{2\pi i\theta }}\ln _{q}(\theta )\quad .
\end{equation}
It follows easily from this that 
\be{notinv}
 \ln_{q}(\theta) = \sum\limits_{k=0}^{\infty}
 \frac{(-1)^{k}q^{\frac{k\left(k-1\right) }{2}}e^{2\pi i\theta k}}
 {\left( q\right )_{k}}\quad \quad 
 \mbox{and} \quad\quad \frac{1}{\ln _{q}(\theta )}
 =\sum\limits_{k=0}^{\infty }
 \frac{e^{2\pi i\theta k}}{\left( q\right) _{k}}\quad .  
\end{equation}
For explicit calculations it will further turn out to be convenient 
to employ the notations (in which we will omit the explicit 
$q$-dependence as long as $q$ is not varying) 
\be{defx}
 \{ x \}_{y}^{-} := \ln_{q^{y}}( x \tau ) 
 \quad \quad  \mbox{and\quad \quad }
 \{ x \}_{y}^{+} :=
 \ln_{q^{y}}\left( x \tau + 1/2 \right) \, ,
 \quad \quad 0 < x \leq y  \, .
\ee
These blocks have the following obvious properties 
\begin{eqnarray}
 \{ x \}_{y}^{+} \{ x \}_{y}^{-} \=
 \{ 2x \} _{2y}^{-} \ ,&\quad &
 \{ x \} _{y}^{\pm} \=
 \prod_{k=0}^{n-1}  \{ x+ ky \}_{ny}^{\pm} \, ,
 \label{pr1} \\
 \ln_{-q^y} (x\tau) \= \{ x \} _{2y}^- \{ x+y \} _{2y}^+ \ , &\quad &
 \ln_{-q^y} (x\tau+1/2) \= \{ x \} _{2y}^+ \{ x+y \} _{2y}^- \, ,  
   \label{pr2} \\
 \{ x \}_{2x}^{-} \= \frac{1}{ \{ x \}_{x}^{+} } \ , &\quad &
 \{ x \}_{2x}^{+} \=  \frac{1}{ \{ x \}_{2x}^{-} \{ 2x \}_{2x}^{+}} \ .
 \label{pr3}
\end{eqnarray}
The last line is Euler's identity which, in fact, can be
derived from (\ref{pr1}).\footnote{Indeed, using consequently the
first and second relation in (\ref{pr1}) for $y\,$=$\,x$, we obtain
$\{ 2x \}_{2x}^{-}=\{ x \}_{x}^{+} \{ x \}_{x}^{-} =
\{ x \}_{x}^{+} \{ x \}_{2x}^{-} \{ 2x \}_{2x}^{-}$, thus deriving
the identity $\{ x \}_{x}^{+} \{ x \}_{2x}^{-} =1$. }

\subsection{Gau{\ss}-Jacobi and Watson Identities} \

\noindent
It will be the principal  aim of our manuscript to seek factorizations 
of some single characters and some linear combinations of characters 
in the following form:
\be{F}
 q^{\rm const} \, \frac{1}{(q)_\infty}\,
 \prod\nolimits_{i=1}^N  \left\{ x_{i}\right\}_{y}^{-} 
 \prod\nolimits_{j=1}^M  \left\{\tilde{x}_{j}\right\}_{y}^{+}  \, .
\end{equation}
We will encounter the cases 
$N\neq 0$, $M=0$ and $N\neq 0$, $M\neq 0$. 
The explicit formulae of this type, which we will obtain, are 
based on the Gau{\ss}-Jacobi identity (see e.g.~\cite{Kac})
\be{GJ}
 \sum_{k=-\infty}^{\infty} (-1)^{k} v^{\frac{k(k+1)}{2}}
 w^{\frac{k(k-1)}{2}} \=
 \prod\limits_{k=1}^{\infty}
 (1-v^{k}w^{k-1}) (1-v^{k-1}w^{k}) (1-v^{k}w^{k}) \,,
\ee
and the Watson identity \cite{Wat}
\ba \label{W}
 \sum_{k=-\infty }^{\infty }v^{\frac{3k^{2}+k}{2}}
 w^{3k^{2}}(w^{-2k}-w^{4k+1}) &=&\prod\limits_{k=1}^{\infty}
 (1-v^{k-1}w^{2k-1}) (1-v^{k}w^{2k-1}) (1-v^{k}w^{2k})  \nn \\
 && \phantom{\prod\limits_{k=1}}
 \times (1-v^{2k-1}w^{4k-4})(1-v^{2k-1}w^{4k}) \, . 
\end{eqnarray}
Substituting $v=q^a$, $w=q^b$, we can  
rewrite the identities in terms of the blocks (\ref{defx})
\ba 
 \label{GJ1}   \sum_{k=-\infty}^{\infty} (-1)^{k} 
  q^{(a+b)\frac{k^2}{2} + \frac{k}{2}(a-b)} &=&
  \{ a \}_{a+b}^{-} \{ b \}_{a+b}^{-} \{ a+b \}_{a+b}^{-} \, ,  \\
  \nn   \sum_{k=-\infty }^{\infty }  q^{\frac{3k^2}{2}(a+2b)} 
 \Bigl(q^{k(a/2-2b)} - q^{k(a/2+4b)+b} \Bigr)  &=& 
 \{ b \}_{a+2b}^{-} \{ a+b \}_{a+2b}^{-} \{ a+2b \}_{a+2b}^{-} \\
 && \times \{ a \}_{2a+4b}^{-} \{ a+4b \}_{2a+4b}^{-} \, . \label{W1} 
\ea
Other useful substitutions are
$v=q^a$, $w=-q^b$ and $v=-q^a$, $w=q^b$ (for (\ref{GJ}) it suffices
to consider only the first of them, because of the symmetry 
$v\leftrightarrow w$), which yield
\ba 
 \label{GJ2}
 && \sum_{k=-\infty}^{\infty} (-1)^{\frac{k(k+1)}{2}} 
  q^{(a+b)\frac{k^2}{2} + \frac{k}{2}(a-b)}  \=  \nn \\ [1mm]
 && \{a\}_{2(a+b)}^{-}\{b\}_{2(a+b)}^{+}\{a+b\}_{2(a+b)}^{+}
 \{a+2b\}_{2(a+b)}^{-}\{2a+b\}_{2(a+b)}^{+}\{2a+2b\}_{2(a+b)}^{-} 
 \, , \\ [1mm]  && \sum_{k=-\infty }^{\infty } (-1)^{3k^2} 
 q^{\frac{3k^2}{2}(a+2b)} (q^{k(a/2-2b)} + q^{k(a/2+4b)+b}) 
  \= \nn \\ [1mm]
 && \{ b \}_{a+2b}^+ \{ a+b \}_{a+2b}^+ \{ a+2b \}_{a+2b}^{-}  
 \{ a\}_{2a+4b}^{-} \{ a+4b \}_{2a+4b}^{-} \, , \label{W2} \\ [1mm]
 &&  \sum_{k=-\infty }^{\infty } (-1)^{\frac{k(k-1)}{2}} 
 q^{\frac{3k^2}{2}(a+2b)} (q^{k(a/2-2b)} - q^{k(a/2+4b)+b}) 
 \=  \{ a \}_{2a+4b}^{+} \{ b \}_{2a+4b}^- \{ a+b \}_{2a+4b}^+ 
 \nn \\ [1mm]  && \times \{ a+2b \}_{2a+4b}^+ \{ a+3b \}_{2a+4b}^+ 
 \{ a+4b \}_{2a+4b}^{+}  \{ 2a+3b \}_{2a+4b}^{-}  
 \{ 2a+4b \}_{2a+4b}^{-}    \, . \label{W3} 
\end{eqnarray}
Here we used (\ref{pr2}) in order to obtain the r.h.s.~in
the desired form.

Now one can try to find factorizable linear combinations
of characters simply by matching the l.h.s.~of (\ref{GJ1})-(\ref{W3})
with appropriate combinations of (\ref{char}). However, this is
a cumbersome task. Below we will develop a more systematic and more 
elegant approach exploiting the quasi-classical asymptotics of 
characters.

\subsection{Quasi-classical  Asymptotics of Characters} \

\noindent 
As can be seen from (\ref{quprod}), the limit $\tau \rightarrow 0$ 
of $\ln_{q}(\theta )$ (since we require ${\rm Im}(\tau )>0$, 
this is the limit $q\rightarrow 1^-$) is singular.
The asymptotics is given by
\be{dilim}
 \lim\limits_{\tau \rightarrow 0} \ln_{q}(\theta )
 \= \exp \left\{ \frac{1}{2\pi i\tau }
 \; Li_{2} \left( e^{2\pi i\theta }\right) +
 \frac 12 \ln(1-e^{2\pi i\theta }) + {\cal O}(\tau)
 \right\} \, ,  
\ee
where $Li_{2} (x)=\sum_{n=1}^{\infty }\frac{x^{n}}{n^{2}}$ is
the Euler dilogarithm 
\footnote{This motivated the authors of \cite{FK} to coin 
 $\ln_{q}(\theta )$ a quantum dilogarithm.}
(see e.g.~\cite{Lewin}). 

Introducing $\hat{q}=\exp\{-2\pi i/\tau\}$, we derive from 
(\ref{defx}) and (\ref{dilim}) the following asymptotics for 
the limit $q \rightarrow 1^-$
\be{xlim}
 \{ x \}_{y}^{-}
 \,\sim\, \hat{q}^{\frac{Li_{2}(1)}{4\pi^2 y}} 
 \= \hat{q}^{\,\frac{1}{24 y}} \ , \quad \quad\quad
  \{ x \}_{y}^{+}
 \,\sim\, \hat{q}^{\frac{Li_{2}(-1)}{4\pi^2 y}} 
 \= \hat{q}^{-\frac{1}{48 y}}  \, .
\ee
Here we used the fact that $Li_{2}(1)=-2 Li_{2}(-1) = \pi^2/6$ 
holds.\footnote{Eqs.~(\ref{xlim}) can also be obtained by
a saddle point analysis of the identities (\ref{notinv})
for $\ln_{q^b}(\theta)$ if we put $\theta=x\tau + 1/2$ and 
$\theta=x\tau$, respectively \cite{BF1}. In this approach one
finds: 
$ \{ x \}_{y}^{-} \sim \hat{q}^{\frac{L(1)}{4\pi^2 y}}$,  
$ \{ x \}_{y}^{+} \sim \hat{q}^{-\frac{L(1/2)}{4\pi^2 y}}$,
as $\tau \rightarrow 0$.
Here $L(z)=Li_2(z)+\frac{1}{2}\ln z\ln (1-z)$ 
denotes the Rogers dilogarithm \cite{Lewin}. These results
coincide with (\ref{xlim}) since  
$L(1)=2 L(1/2) = \pi^2/6$.           }
Notice that $q\rightarrow 1^-$ implies that $\hat{q}\rightarrow 0^+$,
so that $\{ x \}_{y}^{-}$ and $\{ x \}_{y}^{+}$
tend to zero and infinity, respectively.

{}From a physical point of view, say if we regard $\chi(q)$ as a
partition function, the limit $\tau \rightarrow 0$ can be interpreted 
as a high-temperature limit (with temperature $T\sim 1/\tau$)
which is singular and known to be ruled by the effective central 
charge only (i.e.~it is sector-independent) \cite{ceff}. 
Indeed, in order to carry out this limit, one may exploit
the behaviour of Virasoro characters under the modular transformation. 
It is well known \cite{itz,cap}, that the S-modular transformation 
($q\leftrightarrow \hat{q}$) of a character has the following form:  
\be{str}
 \chi_{n,m}^{s,t}(q) \= 
 \sum\limits_{n^{\prime },m^{\prime }}S_{nm}^{n^{\prime}m^{\prime }}
 \chi_{n^{\prime},m^{\prime }}^{s,t}(\hat{q}) \, ,
\end{equation}
where $S_{nm}^{n^{\prime }m^{\prime }}$ are explicitly known 
constants (see (\ref{Sm})). 
Now it is obvious from (\ref{char}) and (\ref{str}) that 
\be{q1}
   \chi_{n,m}^{s,t}(q) \,\sim\,  S_{nm}^{\bar{n}\bar{m}} \, 
 \hat{q}^{\,-\frac{c_{\rm eff}(s,t)}{24}} \quad \
  ( q \rightarrow 1^- ) \,  .  
\ee
Here we have introduced the so-called effective central charge 
$ c_{\rm eff}(s,t) = c(s,t) -24 h^{s,t}_{\bar{n},\bar{m}} =
 1-\frac{6}{st}(\bar{n}t-\bar{m}s)^{2}$,
where $h^{s,t}_{\bar{n},\bar{m}}$ 
denotes the lowest of all conformal weights in the model. 
Let us remark that the conditions (\ref{st}) and (\ref{nm})
allow us to invoke the well-known theorem of the greatest 
common divisor and show that $|\bar{n}t-\bar{m}s| = 1$.
Hence 
\be{cef}
 c_{\rm eff}(s,t) \=  1-\frac{6}{s\,t} 
\end{equation}
holds for any minimal model.

Comparison of (\ref{q1}) with (\ref{xlim}) imposes a  constraint
on the possible structure of characters factorized in form (\ref{F}).
Namely, each factor of the type 
$(\{ x \}_{y}^{-})^{\pm 1}$ and $(\{ x \}_{y}^{+})^{\pm 1}$
contributes,  $\mp\frac 1y$ and $\pm\frac{1}{2y}$ to
the effective central charge, respectively. Notice that this is an
$x$  independent property. 
These contributions must sum up to the value given by (\ref{cef}).

\section{Factorization of Characters } \

\noindent
Below it will be useful to refer to the following simple statement:
\be{zet}
 \zeta \, nm + \zeta^{-1} st \= nt +  ms \ \ \Longleftrightarrow
 \ \ t = \zeta m \ \ {\rm or}\ \ s = \zeta n \, .
\ee  
Equations of these form will arise as necessary conditions for
factorization of (combinations of) characters.
Clearly, for $s$, $t$, $n$, $m$ obeying (\ref{st}) and (\ref{nm})
the parameter $\zeta$ may assume only some rational values 
greater than unity. 

\subsection{Factorization of single Characters} \ 

\noindent
The factorization of some Virasoro characters in the ${\cal M}(3,4)$ 
and ${\cal M}(4,5)$ models was already observed  in \cite{Ro}, 
whereas the factorization of all characters of type 
$\chi_{n,m}^{2n,t}(q)$ and $\chi_{n,m}^{3n,t}(q)$ was discovered in 
\cite{chr}. It was already discussed in \cite{cap,chr,KRV,ES,BF1} 
that the factorization of characters in these series may be obtained 
by exploiting the Gau{\ss}-Jacobi and Watson identities. Nevertheless, 
we wish to present here a systematic derivation of these results 
based on alternative arguments which 
will also be applicable in a more general situation.

It is straightforward to see from (\ref{char}) that the first
three terms in the expansion of the incomplete character are
\be{exp}
\hat{\chi}_{n,m}^{s,t}(q) \=  
 1 - q^{nm} - q^{(s-n)(t-m)} + \ldots  \,,
\ee
and that further terms are of higher powers in $q$.
Let us assume that the incomplete character in question 
is a particular case of the l.h.s.~of the Gau{\ss}-Jacobi identity
(\ref{GJ1}) for some $a$ and $b$. Noticing that the series on the 
l.h.s.~of (\ref{GJ1}) is $1-q^a-q^b +$ higher order terms,
we conclude that $a=nm$, $b=(s-n)(t-m)$ or vice versa.
Furthermore, the r.h.s.~of (\ref{GJ1}) allows to calculate
the effective central charge for the character in question.
As was explained in subsection 1.4, each of the three blocks 
contributes $-\frac{1}{a+b}$ to $c_{\rm eff}$.
Therefore, $c_{\rm eff}=1-\frac{3}{a+b}$ (the 1 is a contribution
of $(q)_{\infty}=\ln _{q}(\tau )=\{ 1 \}_{1}^{-}$, which appears in 
(\ref{char}) 
and whose limit is also ruled by (\ref{xlim})). 
Comparison of this result with (\ref{cef}) yields the equation
\be{z2}
 2 \, nm + \frac 12 \,st \= nt +  ms \, ,
\ee
which is a particular case of (\ref{zet}) with $\zeta=2$ and, hence,
either $s=2n$ or $t=2m$. This implies that $\chi_{n,m}^{2n,t}(q)$ 
is the only (up to the symmetries (\ref{hsym})) possible type of 
characters factorizable with the help of the Gau{\ss}-Jacobi identity 
and that its factorization has to be of the following form:
\be{2sing}
 \chi_{n,m}^{2n,t}(q) \=
 \frac{\,q^{\, h_{n,m}^{2n,t} -\frac{c(2n,t)}{24}} }{(q)_{\infty }}
 \{nm\}_{nt}^{-} \{nt-nm\}_{nt}^{-} \{nt\}_{nt}^{-} \, ,
\ee
where $t$ is an odd number according to (\ref{st}).
One can verify that eq.~(\ref{2sing}) is indeed valid by a direct
matching of the l.h.s.~of (\ref{GJ1}) for the specified $a$ and $b$
with the formula (\ref{char}) for characters (see e.g.~\cite{BF1}).

The same type of consideration applies if we seek characters which
are factorizable with the help of the Watson identity.
Namely, since the series on the l.h.s.~of (\ref{W1}) is again
$1-q^a-q^b +$ higher order terms, we conclude that 
$a=nm$, $b=(s-n)(t-m)$ or vice versa
(in contrast to the previous case, these two possibilities lead
to different equations). 
The r.h.s.~of (\ref{W1}) allows to calculate 
the effective central charge: $c_{\rm eff} = 1-\frac{4}{y}$, 
where $y=a+2b$.
Comparison of this value for 
$c_{\rm eff}$ for the two choices of $a$ and $b$
with (\ref{cef}) yields the following equations
\be{z3}
 \frac 32 \, nm + \frac 23 \,st \= nt +  ms \, \quad \mbox{and} \, 
 \quad 3 \, nm + \frac 13 \,st \= nt +  ms \, ,
\ee
respectively. According to (\ref{zet}) this implies:
$n=2 s/3$ or $m= 2 t/3$ in the first case and 
$n=s/3 $ or $m=t/3 $ in the second. Notice that these
cases are related via the symmetries (\ref{hsym}).
Thus, we conclude that the only possible type of characters 
factorizable on the base of the Watson identity is 
$\chi_{n,m}^{3n,t}(q)$ (again up to the symmetries (\ref{hsym}))
and that its factorization has to be of the following form:
\begin{eqnarray}
 \chi_{n,m}^{3n,t}(q) &=&
 \frac{q^{\,h_{n,m}^{3n,t}-\frac{c(3n,t)}{24} }}{(q)_{\infty }}
  \{nm\}_{2nt}^{-}  \{2nt-nm\}_{2nt}^{-} \{2nt\}_{2nt}^{-}
 \nonumber \\ &&\times 
 \{2nt-2nm\}_{4nt}^{-} \{2nt+2nm\}_{4nt}^{-} \, , \label{3sing} 
\end{eqnarray}
where $\langle 3,t \rangle = 1$.
Again, one verifies this formula directly matching it with
(\ref{char}) (see \cite{BF1}).

Thus, we have found all types of characters which are factorizable
on the base of the Gau{\ss}-Jacobi and the Watson identities.
In fact, it was shown in \cite{chr} that this exhausts the list
of characters of minimal Virasoro models which admit the 
form (\ref{F}) with M=0 and $x_i\neq x_k$.
This implies that for the purpose of factorizing a single character 
in such a form one does not have to invoke the higher Macdonald 
identities \cite{macd} 
(also known as the Weyl-Macdonald denominator identities).

As a last remark in this subsection, we notice that
in the case 
$\langle 2,m \rangle = \langle 3,n \rangle = \langle n,m \rangle = 1$
the combination of (\ref{2sing}) and (\ref{3sing}) yields 
\be{23sing}
 \chi_{n,m}^{2n,3m}(q) \=
 \frac{q^{\,\frac{nm-1}{24}}}{(q)_\infty} \{nm\}^-_{nm}  \, .   
\ee
The first non-trivial example of this kind is 
$\chi_{1,2}^{3,4}(q) =  q^{\frac{1}{24}}/ \{1\}_{2}^{-} 
= q^{\frac{1}{24}}  \{1\}_{1}^{+}$ (the second equality is 
due to the Euler identity). Furthermore, noticing
the symmetry $n \leftrightarrow m$ of the r.h.s.~of eq.~(\ref{23sing}), 
we derive an identity relating different models (it can also be
obtained employing (\ref{alp}) twice)
\be{34sing}
 \chi_{n,m}^{2n,3m}(q) \=   \chi_{m,n}^{2m,3n}(q) \, ,
\ee
where
$\langle 6,n \rangle = \langle 6,m \rangle = \langle n,m \rangle = 1$.
The first non-trivial example is 
$\chi_{1,5}^{2,15}(q) \=   \chi_{1,5}^{3,10}(q)$.

\subsection{Factorization of linear Combinations. 
 Preliminary Ideas} \ 

\noindent
We commence the investigation of factorized linear combinations, 
$\chi_{n,m}^{s,t}(q)\pm \chi_{n',m'}^{s,t}(q)$, by
introducing the quantity
\be{hh}
 \D h_{n,m}^{n',m'}(s,t) :=  h^{s,t}_{n',m'} - h^{s,t}_{n,m} \= 
 \frac{\left( (m+m')s - (n+n')t \right)
 \left((n-n')t-(m-m')s\right)}{4\;s\;t} \, ,
\ee
where we will often omit the labels $s$ and $t$.
Then 
\be{dchr}
\chi_{n,m}^{s,t}(q)\pm \chi_{n',m'}^{s,t}(q) \=
 \frac{ q^{h_{n,m}^{s,t} - \frac{c(s,t)}{24}} }{ (q)_\infty } 
 \left( \hat{\chi}_{n,m}^{s,t}(q) \pm q^{\, \D h_{n,m}^{n',m'} }
 \hat{\chi}_{n',m'}^{s,t}(q) \right) \, .
\ee  
The combination 
 $\hat{\chi}_{n,m}^{s,t}(q) \pm q^{\, \D h_{n,m}^{n',m'} }
 \hat{\chi}_{n',m'}^{s,t}(q)$ can be represented 
as a product of few blocks $\{\phantom{x}\}^\pm$ only if 
$ \D h_{n,m}^{n',m'}$ is an integer or a fraction with
sufficiently small denominator
(other\-wise the product will generate terms with powers of 
$q^{ \D h_{n,m}^{n',m'} }$ which are not presented in 
the combination). On the other hand, the
numerator in (\ref{hh}) is, in general, not divisible by
$st$ because of the conditions (\ref{st}) and (\ref{nm}). 
The only possibility to make this fraction reducible by $st$
is to put $n=n'$ and $m+m'=t$ or, alternatively, $m=m'$ and $n+n'=s$.
Thus, we are led to consider the combinations
\be{com}
 \chi_{n,m}^{s,t}(q)\pm \chi_{n,t-m}^{s,t}(q) \, .
\ee 
Let us denote $\D h_{n,m}^{n,t-m}(s,t)$ for such pairs by
$\D h_{n,m}^{s,t}$; its explicit value is
\be{dhp}
 \D h_{n,m}^{s,t} \= \frac 14 (t-2m)(s-2n) \, .
\ee

If $s$ or $t$ is even (in particular, this includes all 
unitary minimal models), then $\D h_{n,m}^{s,t}$ 
is integer or semi-integer.
Let, for definiteness, $s$ be even. Then, taking into account
the symmetries (\ref{hsym}), we see that each character in the minimal
model ${\cal M}(s,t)$ is either of the form $\chi_{s/2,m}^{s,t}(q)$
(i.e.,~a ``single'' character, factorizable per se)
or there exists exactly one more character such that they form
a pair of type (\ref{com}). 
It follows from this and eq.~(\ref{nm}) that 
the model ${\cal M}(s,t)$ has ${\cal D}_0 = \frac{t-1}{2}$ ``single'' 
characters. Consequently, there are 
${\cal D}_1 = ({\cal D} - {\cal D}_0)/2 = \frac{(s-2)(t-1)}{4}$ pairs.
If both $s$ and $t$ are odd, then apparently ${\cal D}_0 = 0$
and ${\cal D}_1 = {\cal D}/2 = \frac{(s-1)(t-1)}{4}$.

Consider (\ref{com}) for $n$ and $m$ in the range
\be{nmst}
  n < s/2 \ , \quad \quad m < t/2 \, .
\ee
For this range all involved characters are different (see 
e.g.~\cite{cap}), and it is easy to see that we cover all 
possible ${\cal D}_1$ combinations.
Moreover, conditions (\ref{nmst}) ensure that 
$\D h_{n,m}^{s,t} > 0 $. This in turn implies that (\ref{com})  
contains only non-negative powers of $q$. 
Thus, from now on we will assume that $n$ and $m$ in 
(\ref{com}) are restricted as in (\ref{nmst}).

As we have seen in the previous subsection, the knowledge of the 
asymptotic  behaviour of the characters in the $q\rightarrow 1^-$ limit
proves to be very useful in the search of factorized characters.
It turns out that in the case of linear combinations we have
to take into account also the next to leading term in (\ref{q1}),
\be{lim}
 \chi_{n,m}^{s,t}(q) \ \sim \   S_{nm}^{\bar{n}\bar{m}} \, 
 \hat{q}^{\,-\frac{c_{\rm eff}(s,t)}{24}} +
 S_{nm}^{\tilde{n}\tilde{m} } \,
 \hat{q}^{\,-\frac{\tilde{c}(s,t)}{24}} + \ldots \quad \
 (  q \rightarrow 1^-  ) \, ,
\ee
where we denoted
\be{efff}
c_{\rm eff}(s,t) \= c(s,t)-24\,h^{s,t}_{\bar{n},\bar{m}} \ ,  
 \ \ \ \ \    \tilde{c}(s,t) \= 
 c(s,t)-24\,h^{s,t}_{\tilde{n},\tilde{m}} \ .
\ee
Here $h^{s,t}_{\bar{n},\bar{m}}$ and 
$h^{s,t}_{\tilde{n},\tilde{m}}$ are the smallest and the 
second smallest conformal weights in the model corresponding
to the minimal and the next to minimal value of $|nt-ms|$, 
respectively. We will refer to $\tilde{c}(s,t)$ as the 
secondary effective central charge.

As we mentioned above, the theorem of the greatest common divisor 
ensures that $c_{\rm eff}(s,t) = 1- 6/st$.
Furthermore, one can show in the same way
that $|\tilde{n} t - \tilde{m} s |=2$, so that 
$h^{s,t}_{\tilde{n},\tilde{m}}=\frac{4-(s-t)^{2}}{4\;s\;t}$
holds.%
\footnote{In fact, a more general statement is valid: 
for $s$ and $t$ obeying (\ref{st}) and positive integer $k$ 
such that $k < \min (s,t)$ there exists always a solution of the 
equation $|n t - m s |=k$ obeying (\ref{nm}). It is given by
$n=k\bar{n} - p s$ and  $m=k\bar{m} - p t$, where 
$p$ is some integer depending on $k$.} 
 Thus, employing (\ref{c}) and (\ref{efff}), we obtain
\be{ceff}
 \tilde{c}(s,t) \= 1-\frac{24}{s\,t} \ .
\ee
The only case where this argument fails is ${\cal M}(2,t)$.
Here $|\tilde{n} t - \tilde{m} s |=3$ (unless $t=3$, in which case 
$h^{s,t}_{\tilde{n},\tilde{m}}$ does not exist).
But, as we demonstrated, in this case all the characters are
factorizable per se.

Now, using the explicit form of the matrix $S$  \cite{itz,cap} 
involved in the S-modular transformation (\ref{str})
\be{Sm}
 S_{n,m}^{n^{\prime},m^{\prime }} \=
  \sqrt{\frac{8}{st}}\, (-1)^{nm'+mn'+1} \, 
  \sin\Bigl(\frac{\pi \, n n' t}{s}  \Bigr)
  \sin\Bigl(\frac{\pi \, m m' s}{t}  \Bigr) \, ,
\ee
we observe that 
$S_{n,t-m}^{n',m'}=-S_{n,m}^{n',m'} (-1)^{n' t - m' s}$.
Taking into account that $|\bar{n} t - \bar{m} s|=1$ and
$|\tilde{n} t - \tilde{m} s|=2$, we conclude that 
$S_{n,t-m}^{\bar{n},\bar{m}}=S_{n,m}^{\bar{n},\bar{m}} $ and
$S_{n,t-m}^{\tilde{n},\tilde{m}}=-S_{n,m}^{\tilde{n},\tilde{m}}$.
Therefore, for the combination
$\chi_{n,m}^{s,t}(q) - \chi_{n,t-m}^{s,t}(q)$ 
the leading terms  in (\ref{lim}) corresponding to $c_{\rm eff}(s,t)$ 
cancel but those corresponding to $\tilde{c}(s,t)$ survive%
\footnote{For ${\cal M}(2,t)$ these terms cancel since
$|\tilde{n} t - \tilde{m} s|=3$. This is not surprising because
in this case $\chi_{1,m}^{2,t}(q) - \chi_{1,t-m}^{2,t}(q)$
vanishes due to (\ref{hsym}).} 
and for the combination 
$\chi_{n,m}^{s,t}(q) + \chi_{n,t-m}^{s,t}(q)$ 
the leading terms corresponding to $c_{\rm eff}(s,t)$ do not cancel.
Thus, we obtain the following asymptotics of the combinations
in question
\begin{eqnarray}
 && \chi_{n,m}^{s,t}(q) + \chi_{n,t-m}^{s,t}(q)  
 \ \sim\   \hat{q}^{\,-\frac{c_{\rm eff}(s,t)}{24}} \
 \quad ( q \rightarrow 1^- )   \, , \label{q1a} \\
  &&  \chi_{n,m}^{s,t}(q) - \chi_{n,t-m}^{s,t}(q) 
  \ \sim\   \hat{q}^{\,-\frac{\tilde{c}(s,t)}{24}} \quad
 \quad (q \rightarrow 1^- ) \,  .  \label{q1b}
\end{eqnarray}

\subsection{Factorization of linear Combinations. Exact Formulae} \

\noindent
Now we are in the position to find all combinations of type 
(\ref{com}) which are factorizable on the base of the Gau{\ss}-Jacobi 
and Watson identities. First, it follows from (\ref{char}) that 
\be{exp2}
 \hat{\chi}_{n,m}^{s,t}(q) \pm q^{\D h_{n,m}^{s,t}}
  \hat{\chi}_{n,t-m}^{s,t}(q)\=  
 1 - q^{nm} \pm q^{ \D h_{n,m}^{s,t} } + \ldots 
\ee
and the further terms are of higher powers in $q$. Here
$\D h_{n,m}^{s,t} $ is given by (\ref{dhp}) and we assume
$n < s/2$, $m < t/2$, notice that then $nm \neq \Delta h^{s,t}_{n,m}$

We will consider the sum of characters first. Let us
assume that it is factorizable on the base of the Gau{\ss}-Jacobi 
identity (\ref{GJ2}), whose expansion on the l.h.s.~is
$1 - q^a + q^b +$ higher order terms. Then we infer from (\ref{exp2})  
that $a=nm$ and $b= \D h_{n,m}^{s,t} $.
The r.h.s.~of (\ref{GJ2}) gives the following effective
central charge of the combination in question: 
$c_{\rm eff}=1-\frac{3}{4(a+b)}$. Comparing it with (\ref{cef}),
we obtain the equation $8(a+b)=st$, or more explicitly
\be{z4}
 4 \, nm + \frac 14 \,st \= nt +  ms \, .
\ee
According to (\ref{zet}), this implies $4 n=s$ or 
$4 m=t $. 

If we assume that the difference of characters in (\ref{exp2}) 
is factorizable on the base of the Gau{\ss}-Jacobi identity 
(\ref{GJ1}), we have to put $a=nm$, $b=\D h_{n,m}^{s,t}$ or vice versa. 
According to (\ref{q1b}), the asymptotics $q\rightarrow 1^-$ defines 
the secondary effective 
central charge, and comparison with the r.h.s. of (\ref{GJ1}) yields
$\tilde{c} = 1- \frac{3}{a+b}$. Together with (\ref{ceff}),
we obtain the equation $8(a+b)=st$ which leads to the same
condition (\ref{z4}) found for the sum of characters. 

Thus, we have shown that the only possible 
(up to the symmetries (\ref{hsym})) combination of 
characters factorizable on the base of the Gau{\ss}-Jacobi identity is 
$\chi_{n,m}^{4n,t}(q) \pm \chi_{n,t-m}^{4n,t}(q)$
and that its factorization has to be of the following form 
\ba
 \chi_{n,m}^{4n,t} (q) + \chi_{n,t-m}^{4n,t} (q) &=&
 \frac{q^{h^{4n,t}_{n,m}-\frac{c(4n,t)}{24}}}{(q)_{\infty }}
 \{ nm \}_{nt}^-  \{ nt-nm \}_{nt}^- \{ nt \}_{nt}^- \nonumber \\
 &&\times \{ nt/2 - nm \}_{nt}^{+}  \{ nt/2  \}_{nt}^{+}
  \{ nt/2 +nm \} _{nt}^{+} \, , \label{4p} \\ [1mm]
 \chi_{n,m}^{4n,t} (q) - \chi_{n,t-m}^{4n,t} (q) &=&
 \frac{q^{h^{4n,t}_{n,m}-\frac{c(4n,t)}{24}}}{(q)_{\infty }} 
 \{ nm \}_{\frac{nt}{2}}^{-}
 \{ nt/2 - nm \}_{\frac{nt}{2}}^{-}
 \{ nt/2 \}_{\frac{nt}{2}}^{-} \, . \label{4m}
\ea 
Here $\langle t,2 \rangle=\langle t,n \rangle=1$. 
The direct proof of these relations is performed again by 
matching them with (\ref{char}) (see appendix B). 
Notice that in the case of odd $n$ it suffices to prove only one
of the relations, say (\ref{4m}). Indeed, in this case
$\D h_{n,m}^{4n,t} = n(t-2m)/2$ is semi-integer, so that
changing the signs of all semi-integer powers in the series
on the l.h.s.~of (\ref{4p}), we obtain the series on the l.h.s.~of 
(\ref{4m}). Therefore, the r.h.s.~of (\ref{4p}) is derived from 
the r.h.s.~of (\ref{4m}) with the help of (\ref{pr2}).

Now we apply the same technique as above in order to find
the differences of type (\ref{com})
which are factorizable on the base of the Watson identity. 
We assume they have the form
of eq.~(\ref{W1}), whose expansion on the l.h.s.~is
$1 - q^a - q^b +$ higher order terms. Then we infer from 
(\ref{exp2}) that $a=nm$ and $b= \D h_{n,m}^{s,t}$ or 
$a = \D h_{n,m}^{s,t}$ and $b=nm$.
The r.h.s.~of (\ref{W1}) gives the following secondary 
effective central charge of the combination in question: 
$\tilde{c}=1-\frac{4}{a+2b}$. Comparing it with (\ref{ceff}),
we obtain the equation $6(a+2b)=st$, which gives for
the two possible choices of $a$ and $b$
\be{z5}
 3 \, nm + \frac 13 \,st \= nt +  ms \, , \quad
 {\rm and} \quad 
 6 \, nm + \frac 16 \,st \= nt +  ms \, ,
\ee
respectively. According to (\ref{zet}), this implies 
$3 n=s $ or $3 m=t $ in the first case and  
$6 n=s $ or $6 m=t $ in the second.

Assuming that the sum of characters in (\ref{exp2}) 
is factorized on the base of the Watson identity (\ref{W2}), 
we have to put $a=nm$, $b=\D h_{n,m}^{s,t}$. Then 
the r.h.s.~of (\ref{W2}) gives $c_{\rm eff}=1-\frac{1}{a+2b}$.
Comparing it with (\ref{cef}), we obtain the equation $6(a+2b)=st$ 
and, thus, we recover the first equation in (\ref{z5}).
So, this is once more the case $n=s/3 $ or $m=t/ 3 $.

It turns out that the sum of characters in (\ref{exp2}) 
cannot be factorized on the base of the Watson identity (\ref{W3}).
Indeed, its l.h.s.~is the following series
$ 1 + q^a - q^b + q^{2a+b} - q^{a+5b} +$ higher order
terms. On the other hand, for $n < s/2$, $m < t/2$ we have
\be{exp3}
 \hat{\chi}_{n,m}^{s,t}(q) \pm q^{ \D h_{n,m}^{s,t} }
  \hat{\chi}_{n,t-m}^{s,t}(q)\=  
 1 - q^{nm} \pm q^{ \D h_{n,m}^{s,t} } 
 \mp q^{\Delta h_{n,m}^{s,t} + n(t-m) } 
 \mp q^{\Delta h_{n,m}^{s,t} + m(s-n) } + \ldots  ,
\ee
where further terms are of higher powers in $q$. 
Evidently, these two series cannot match because of the
wrong sign of the $q^{2a+b}$ term.

Thus, the only possible 
(up to the symmetries (\ref{hsym})) combinations of 
characters factorizable on the base of the Watson identity are
$\chi_{n,m}^{3n,t}(q) \pm \chi_{n,t-m}^{3n,t}(q)$ and 
$\chi_{n,m}^{6n,t}(q) - \chi_{n,t-m}^{6n,t}(q)$ and  
their factorizations have to be of the following form
\begin{eqnarray}   
 \label{3sum} \chi _{n,m}^{3n,t} (q) \pm \chi_{n,t-m}^{3n,t} (q)
 &=&  \frac{q^{h_{n,m}^{3n,t}-\frac{c(3n,t)}{24} }}{(q)_\infty }
 \{ nm \}_{nt}^- \{ nt-nm \}_{nt}^- \nonumber \\
 &&\times  \left\{ \frac{nt}{2} \right\}_{\frac{nt}{2}}^-
 \left\{ \frac{nt-2nm}{4}\right\}_{\frac{nt}{2}}^{\pm }  
 \left\{ \frac{nt+2nm}{4}\right\}_{\frac{nt}{2}}^{\pm } \, ,  \\
 \label{6sum}   \chi_{n,m}^{6n,t} (q) - \chi_{n,t-m}^{6n,t} (q)
 &=& \frac{q^{h_{n,m}^{6n,t} - \frac{c}{24}(6n,t) }}{(q)_\infty }
 \{ nm \}_{nt}^- \{ nt-nm \}_{nt}^- \{ nt \}_{nt}^- \nonumber \\
 &&\times  \{ nt-2nm \}_{2nt}^-  \{ nt+2nm \}_{2nt}^- \, . 
 \end{eqnarray}
Here $\langle t,3 \rangle=\langle t,n \rangle =1$ in (\ref{3sum}) 
and $\langle t,6 \rangle=\langle t,n \rangle=1$ in (\ref{6sum}).
The direct proof of these relations is performed again by matching 
them with (\ref{char}) (see appendix B).

Combining (\ref{4p})-(\ref{4m}) and (\ref{3sum}), we also obtain  
\ba
 \chi_{n,m}^{4n,3m}(q) + \chi_{n,2m}^{4n,3m}(q)
 &=& \frac{q^{\,\frac{nm-2}{48}}}{(q)_\infty} 
  \{nm\}^-_{nm} \{ nm/2 \}^+_{nm} \, , \label{34ta} \\ [1mm]
 \chi_{n,m}^{4n,3m}(q) - \chi_{n,2m}^{4n,3m}(q) &=&
 \frac{q^{\,\frac{nm-2}{48}}}{(q)_\infty} 
 \{ nm/2 \}_{ \frac{nm}{2} }^- \, , \label{34tb}
\ea
where $\langle n,3 \rangle= \langle m,2 \rangle=
 \langle n,m \rangle= 1$.

To conclude this subsection we mention an interesting
byproduct identity, which follows from (\ref{3sing}) and (\ref{3sum}),  
\begin{eqnarray}
 \left\{ nm\right\}_{nt}^{-}\left\{ nt-nm\right\} _{nt}^{-}
 \left\{ \frac{nt}{2}\right\}_{\frac{nt}{2}}^{-} \left\{ 
 \frac{nt-2nm}{4}\right\} _{\frac{nt}{2}}^{\pm } \left\{ 
 \frac{nt+2nm}{4}\right\} _{\frac{nt}{2}}^{\pm } &=& \nonumber  \\
 \left\{ 2nt\right\} _{2nt}^{-}\left\{ nm\right\} _{2nt}^{-}\left\{
 2nt-nm\right\} _{2nt}^{-}\left\{ 2nt-2nm\right\} _{4nt}^{-}\left\{
 2nt+2nm\right\} _{4nt}^{-} && \nonumber  \\
 \pm q^{\frac{nt-2nm}{4}}\left\{ 2nt\right\} _{2nt}^{-}
 \left\{ nt-nm\right\}_{2nt}^{-}\left\{ nt+nm\right\}_{2nt}^{-}
 \left\{ 2nm\right\}_{4nt}^{-}\left\{ 4nt-2nm\right\}_{4nt}^{-} \,, &&  
\end{eqnarray}
and may also be rewritten as 
\begin{eqnarray}
 \left\{ \frac{nt}{2} \right\}^-_{nt} 
  \left\{ \frac{nt-2nm}{4} \right\}^\pm_{\frac{nt}{2}}
  \left\{ \frac{nt+2nm}{4} \right\}^\pm_{\frac{nt}{2}} &=&
  \{ nt \}^+_{nt} \{ nt-nm \}^+_{2nt}\{ nt+nm \}^+_{2nt}   \\ 
&& \pm   q^{ \frac{nt-2nm}{4} } 
  \{ nt \}^+_{nt} \{ nm \}^+_{2nt} \{ 2nt-nm \}^+_{2nt} \, . 
\nonumber
\end{eqnarray}

This identity resembles  particular  formulae in \cite{Tao} 
( (A5)  and (A6) therein ), which were useful to derive a different
type of identities between characters.

Analogous identities following from (\ref{34ta})-(\ref{34tb})
and (\ref{3sum}) are
\be{34id}
 \{3nm\}^+_{8nm} \{5nm\}^+_{8nm} \pm
 q^{ \frac{nm}{2} } \{nm\}^+_{8nm} \{7nm\}^+_{8nm}  
 = \{ nm/2 \}^\pm_{nm} 
 \{2nm\}^-_{4nm} \{4nm\}^-_{8nm} . 
\ee

\subsection{Remarks on the factorized Combinations} \

\noindent
The factorized characters given by (\ref{2sing}) and (\ref{3sing}) 
can be rewritten in the ``inverse product'' form (examples of such 
representation are given in appendix A)
\footnote{Exactly this form was an aim in \cite{chr}.}
\be{P}
    q^{\rm const } \, \frac{1}{\prod_{i=1}^N \{x_i\}^-_{y_i} } \,.
\ee
In order to achieve this, one rewrites $(q)_{\infty }=\{1\}_{1}^{-}$ 
with the help of (\ref{pr1}) as a product of some number of blocks
and then cancels all blocks in the numerator with some of those in 
the denominator. The only problem here is to verify that 
all blocks in (\ref{2sing}) and (\ref{3sing}) are different.
Eq.~(\ref{2sing}) could have coinciding blocks only
if $t=2m$. This is however  excluded by the condition 
$\langle t,2 \rangle=1$ which must hold because of (\ref{st}). 
Eq.~(\ref{3sing}) could have coinciding blocks if $t=3m$, $t=3m/2$ 
or $t=2m$. The first two possibilities are excluded by the condition 
$\langle t,3 \rangle=1$. The last one is allowed, but this case is 
described by the reduced formula (\ref{23sing}), which is obviously 
representable in the form (\ref{P}).

The inverse product form (\ref{P}) (it is rather common for characters
of Kac-Moody algebras \cite{Kac}) can be interpreted as a character of
a module with states created by bosonic type operators. Having the 
characters in the form (\ref{P}) implies that the dimension of the 
level $k$ in the Verma module of the irreducible representation is 
simply the number of partitions 
$k = x_1 + \ldots + x_N + \sum_{i=1}^N n_i y_i $ with $n_i$ 
being an arbitrary non-negative integer. This suggests that the states 
at this level are simply monomials of the form (\ref{states}). If any
power of a generator having a given grading $k$ is allowed, the character 
acquires a factor $(1-q^k)$ in the denominator. It is guaranteed 
that any monomial by itself (apart from $L_{-1}| h=0 \rangle $) can never 
constitute a null-vector, as follows from the following simple argument. 
A null-vector has by definition zero norm or equivalently it is 
annihilated by $L_n$ for all $n>0$. Hence to prove our statement it 
is sufficient to show for one $n$ that  $L_n$ acting on (\ref{states}) 
is non-vanishing. It is easy to verify for $k_1 \neq k $ that $L_k$ 
acting on (\ref{states}) vanishes only for $h=0$. In case 
$ k_1 = k \neq 1$, the action of $L_{k-1}$ is always non-vanishing. 
However, one may not guarantee that all these monomials are linearly 
independent.  

It turns out that all of the factorized combinations of characters 
(\ref{4p})-(\ref{6sum}) and (\ref{34ta})-(\ref{34tb}) can be rewritten
in the inverse product form generalizing (\ref{P}), namely as
\be{PP}
  q^{\rm const } \, \frac{1}{ \prod_{i=1}^N \{x_j\}^-_{y_j} \,
  \prod_{j=1}^M \{\tilde{x}_i\}^+_{y_i}  } \ .
\ee
In particular, (\ref{4m}) for even $n$,
the lower sign in (\ref{3sum}) for integer $nt/4$ and $nm/2$,
and (\ref{6sum}) can be analyzed easily in the way we presented above
and correspond to (\ref{PP}) with $M=0$. The analysis of other
cases is slightly more involved (since we encounter 
$\{\phantom x\}^+$ blocks and blocks with non-integer arguments)
but goes essentially along the same lines. Consider, for instance,
(\ref{4p}). Using (\ref{pr1}) and (\ref{pr3}), we can rewrite its
r.h.s.~as follows (we use here the notation
$\{x_1;\ldots ;x_n\}_y^\pm:=\{x_1\}_y^\pm\ldots \{x_n\}_y^\pm$)
\[
 q^{\rm const} \, \frac{
 \{ nm; n(t - 2m); n(t-m); nt; n(t+m); n(2t-m); n(t + 2m); 2nt \}_{2nt}^-
 }{ \{1\}^-_1 \, \{ nt/2-nm; nt/2; nt/2 +nm) \}_{nt}^- \, 
 \{ nt \}_{nt}^+ } \, .
\]
For $n$ and $m$ in the range (\ref{nmst}) the numerator could
have coinciding blocks only if $t=3m$. However,
in this case we have the reduced formula (\ref{34ta}) which
is readily seen to be representable in form (\ref{PP})
if we take (\ref{pr3}) into account. 
Analysis of (\ref{4m}), (\ref{3sum}) and (\ref{34tb}) 
is performed analogously (notice only that for (\ref{3sum}) one 
has to distinguish the cases $nt/2=0,1\ {\rm mod}\ 2$). 

Thus, all the factorizable combinations of characters of type 
(\ref{com}) admit the form (\ref{PP}). Examples of such 
representation are given in appendix A. Moreover, we prove 
(see appendix C) that there are no other factorizable differences 
of this type which admit the inverse product form (\ref{P}). This 
is a rather surprising fact because the Gau{\ss}-Jacobi and Watson 
identities are the specific Macdonald identities \cite{macd} for 
the $A_1^{(1)}$ and $A_2^{(2)}$ algebras and one could expect that 
the higher Macdonald identities also lead to similar factorizations.

It is worth to notice that some of the factorizable
combinations discussed above admit the following form
\be{PPP}
  q^{\rm const } \, \frac{\prod_{j=1}^M \{\tilde{x}_i\}^+_{y_i} }{ 
  \prod_{i=1}^N \{x_j\}^-_{y_j}    } \ .
\ee
This is the most natural form if we consider such an expression as 
a character (e.g.~in the context of the super-conformal models, see
subsection 3.4) of a module with states created not only by bosonic 
type operators but also by fermionic type operators,
which produce the blocks in the numerator. 
Also, the form (\ref{PPP})
gives particularly simple formulae for quasi-particle momenta 
(see subsection 3.3).

\section{Applications} \ 

\noindent 
In the rest of the paper we will present some corollaries 
and applications of the obtained results
both in a mathematical and physical context. 

\subsection{Identities between Characters} \

\noindent 
We commence by matching the product sides of the formulae for 
the factorized linear combinations of characters with those for
the factorized single characters. For (\ref{4m}) this yields
\be{4t2}
 \chi_{2n,m}^{8n,t} (q) - \chi_{2n,t-m}^{8n,t} (q) \,=\,
  \chi_{n,2m}^{2n,t} (q) \, ,
\ee
where $\langle t, 2 \rangle = \langle t, n \rangle = 1$.
Notice that this identity is exact in the sense that is it does not 
need an extra factor of type $q^{\rm const}$ on the r.h.s.~because
$h_{n,2m}^{2n,t}-c(2n,t)/24=h_{2n,m}^{8n,t}-c(8n,t)/24$. 
~\footnote{This property, which actually holds for all identities
in this subsection, hints on specific modular properties of the
combinations of type (\ref{com}).}
Since $\chi_{1,1}^{2,3}(q)=1$, we obtain, as a particular case, the
identity (which was also presented in \cite{cap} in 
a different context)
\[
  \chi_{1,2}^{3,8}(q) - \chi_{1,6}^{3,8}(q) = 1 \, .
\]
This is the only possible identity of the type
$\chi_{n,m}^{s,t}(q)-\chi_{s,t-m}^{n,t}(q) = q^{\rm const}$ 
because it requires $\tilde{c}(s,t)=0$. According to (\ref{ceff}), 
this implies $st=24$. The latter equation is solved uniquely 
(up to a permutation of $s$ and $t$) due to (\ref{st}).

For (\ref{6sum}) and 
(\ref{3sum}) we obtain analogously 
\be{3t2}
 \chi_{4n,m}^{12n,t} (q) -\chi_{8n,m}^{12n,t} (q) \,=\,
  \chi_{2n,2m}^{3n,t} (q) \, , \quad \
 \chi_{2n,m}^{12n,t} (q) - \chi_{10n,m}^{12n,t} (q) \,=\,
  \chi_{n,2m}^{3n,t} (q) \, ,
\ee
where $\langle t, 6 \rangle = \langle n, t\rangle = 1$. 
These identities are also exact. The first nontrivial examples
of this kind are 
$\chi_{4,m}^{12,5} (q) -\chi_{8,m}^{12,5} (q) =
 \chi_{2,2m}^{3,5} (q)$ and 
$\chi_{2,m}^{12,5} (q) -\chi_{10,m}^{12,5} (q) =
 \chi_{1,2m}^{3,5} (q)$, $m=1,2$.
Furthermore,
the characters on the r.h.s.~of (\ref{3t2}) form a pair of the 
type (\ref{com}), and applying (\ref{3sum}),
we obtain (assuming $m < t/4$ for definiteness)
\begin{eqnarray}    \label{12sum} 
 && \chi_{2n,m}^{12n,t} (q) - \chi_{10n,m}^{12n,t} (q) \pm
\chi_{4n,m}^{12n,t} (q) \mp \chi_{8n,m}^{12n,t} (q)  \= \\
 &&  \frac{q^{h_{n,2m}^{3n,t}-\frac{c(3n,t)}{24} }}{(q)_\infty }
 \{ 2nm \}_{nt}^- \{ nt-2nm \}_{nt}^- 
 \{ nt/2 \}_{\frac{nt}{2}}^- \{ nt/4- nm \}_{\frac{nt}{2}}^{\pm }  
 \{ nt/4 + nm \}_{\frac{nt}{2}}^{\pm } \, .   \nn
\end{eqnarray}
Finally, matching the r.h.s.~of (\ref{12sum}) for the lower 
sign with (\ref{3sing}), we obtain
\be{48f}
 \chi_{8n,m}^{48n,t} (q) - \chi_{16n,m}^{48n,t} (q) 
 + \chi_{32n,m}^{48n,t} (q) - \chi_{40n,m}^{48n,t} (q)  
 \= \chi_{2n,4m}^{3n,t} (q) \, .
\ee
The first nontrivial example is
$\chi_{8,1}^{48,5} (q) - \chi_{16,1}^{48,5} (q) 
 + \chi_{32,1}^{48,5} (q) - \chi_{40,1}^{48,5} (q)  
 = \chi_{1,1}^{3,5} (q)$.

Another way to derive some new identities is to match the 
product sides of different factorized linear combinations.
In particular, one easily recovers the property (\ref{alp}) for 
combinations
\be{alph}
 \chi_{n,\alpha m}^{s,\alpha t}(q) \pm 
 \chi_{n,\alpha(t-m)}^{s,\alpha t}(q) \=
 \chi_{\alpha n,m}^{\alpha s,t}(q) \pm 
 \chi_{\alpha n,t-m}^{\alpha s,t}(q) \,,
\ee
if $\alpha$ is a positive integer such that 
$\langle t, \alpha \rangle = \langle s, \alpha \rangle = 1$.
For instance, 
$\chi_{1,2m}^{3,10}(q) \pm \chi_{2,2m}^{3,10}(q) =
   \chi_{m,2}^{5,6}(q) \pm \chi_{m,4}^{5,6}(q)$, $m=1,2$.

Less obvious identities between characters of different models having 
the same $c_{\rm eff}$ follow if we compare the r.h.s.~of (\ref{3sum}) 
and (\ref{6sum}):
\be{326}
 \chi_{n,t-2m}^{3n,2t}(q) - \chi_{n,t+2m}^{3n,2t}(q) \=
   \chi_{n,m}^{6n,t}(q) - \chi_{5n,m}^{6n,t}(q) \, ,
\ee
where $\langle t, 6 \rangle = \langle n,2 \rangle = 
 \langle t, n \rangle  = 1$, $m<t/2$. For instance,
$ \chi_{1,1}^{3,10}(q) - \chi_{2,1}^{3,10}(q) \=
   \chi_{2,1}^{5,6}(q) - \chi_{2,5}^{5,6}(q)$.

Employing the factorized form of (combinations of) characters,
we can derive identities involving their bilinear combinations.
For instance, it is straightforward to verify the following relations 
(see appendix D for a sample proof)
\begin{eqnarray}
 \chi_{n,m}^{3n,2m} \chi _{2n,m}^{4n,5m} &=&
 \chi_{n,2m}^{3n,4m} \left( \chi_{n,2m}^{6n,5m} -
 \chi_{n,3m}^{6n,5m}\right)  \, , \label{id1} \\
 \chi_{n,m}^{3n,2m} \chi _{2n,2m}^{4n,5m} &=&
 \chi_{n,2m}^{3n,4m} \left( \chi_{n,m}^{6n,5m} - 
 \chi_{n,4m}^{6n,5m} \right) \, , \label{id2} \\
 \chi_{n,m}^{3n,2m} \left(\chi_{n,m}^{4n,5m} \pm
 \chi_{n,4m}^{4n,5m} \right) &=& \left(\chi_{2n,2m}^{6n,5m} \mp  
 \chi_{2n,3m}^{6n,5m} \right) \left(\chi_{n,m}^{3n,4m} \pm 
 \chi_{n,3m}^{3n,4m} \right) \, , \label{id3} \\
 \label{id4}  \chi_{n,m}^{3n,2m}
 \left(\chi_{n,2m}^{4n,5m} \pm \chi_{n,3m}^{4n,5m} \right) &=&
 \left(\chi_{2n,m}^{6n,5m} \mp \chi_{2n,4m}^{6n,5m} \right)
 \left(\chi_{n,m}^{3n,4m} \pm \chi_{n,3m}^{3n,4m} \right) \, , 
\end{eqnarray}
which in turn lead to the identities
\ba
 & \label{id5} \chi_{2n,m}^{4n,5m} \left( \chi_{n,m}^{6n,5m} - 
 \chi_{n,4m}^{6n,5m} \right) \= \chi_{2n,2m}^{4n,5m} 
 \left(\chi_{n,2m}^{6n,5m} - \chi_{n,3m}^{6n,5m} \right) \,, & \\
 & \left( \chi_{n,2m}^{4n,5m} \pm \chi_{n,3m}^{4n,5m} \right)
 \left( \chi_{2n,2m}^{6n,5m} \mp \chi_{2n,2m}^{6n,5m} \right) \= 
 \left( \chi_{n,m}^{4n,5m} \pm \chi_{n,4m}^{4n,5m} \right)
 \left(\chi_{2n,m}^{6n,5m} \mp \chi_{2n,4m}^{6n,5m}\right) \,. &
 \label{id6}
\end{eqnarray}
We have omitted the $q$-dependence for compactness of the formulae.
Once more we like to point out these relations are exact 
(see (\ref{D1})). A particular case of (\ref{id5}) and (\ref{id6}) 
for $n=m=1$ was found in \cite{Tao}. 
Further interesting identities are for instance
\be{3434}
 \left( \chi_{n,m}^{3n,4m}(q) + \chi_{n,3m}^{3n,4m}(q) \right)
  \left( \chi_{n,m}^{3n,4m}(q) - \chi_{n,3m}^{3n,4m}(q) \right) \=
 q^{-\frac{nm}{24}} \left( \chi_{n,m}^{3n,2m}(q) \right)^2 
 \{nm\}^-_{2nm} \,,
\ee
\be{5634}
  \chi_{1,2}^{5,6}(q) \chi_{2,2}^{5,6}(q) -
  \chi_{1,4}^{5,6}(q) \chi_{2,4}^{5,6}(q) \=
  \left( \chi_{1,2}^{3,4}(q) \right)^2 \,,
\ee
\be{4556415a}
  \left( \chi_{1,2}^{4,5}(q) \pm \chi_{1,3}^{4,5}(q) \right)
  \left( \chi_{2,2}^{5,6}(q) \mp \chi_{2,4}^{5,6}(q) \right) \=
  \chi_{1,5}^{4,15}(q) \pm \chi_{3,5}^{4,15}(q) \, .
\ee
Eq.~(\ref{3434}) for $n$=$m$=$1$ yields the well-known relation
\[
\Bigl((\chi_{1,1}^{3,4}(q))^2 - (\chi_{1,3}^{3,4}(q))^2\Bigr)
\chi_{1,2}^{3,4}(q) \=1 \,.
\]

It is of a certain interest to search for relations between 
(combinations of) characters with rescaled $q$. The rescaling, 
$q\rightarrow q^r$ or, equivalently, $\tau\rightarrow r\tau$ can be
regarded as a transformation relating theories on two different tori. 
In statistical mechanics, where $\tau$ is considered as a physical
parameter (e.g.~inverse temperature or width of a strip), this
transformation relates two models at different values of this parameter.

In order to match the factorized (combinations of) characters 
involving those with rescaled $q$ it is useful to take into account 
that such rescaling, $q\rightarrow q^r$, also leads to the rescaling 
of $c_{\rm eff} \rightarrow c_{\rm eff}/r$ (and 
$\tilde{c} \rightarrow \tilde{c}/r$). We present here only several 
examples relating characters of some models with interesting physical 
content under the transformation $q\rightarrow q^2$:
\ba
 &  \chi_{1,1}^{3,4}(q^2) - \chi_{1,3}^{3,4}(q^2) \=
   \left( \chi_{1,2}^{3,4}(q) \right)^{-1}  \, ,  & \\
 & \chi_{1,2}^{5,6}(q^2) + \chi_{1,4}^{5,6}(q^2) \=
   \chi_{1,1}^{2,5}(q) \, , \quad   \label{xxx}
 \chi_{2,2}^{5,6}(q^2) + \chi_{2,4}^{5,6}(q^2) \=
   \chi_{1,2}^{2,5}(q)  \, , & \\
 & \chi_{1,1}^{5,6}(q) - \chi_{1,5}^{5,6}(q) \=
   \chi_{1,2}^{2,5}(q^2) \, , \quad
 \chi_{2,1}^{5,6}(q) - \chi_{2,5}^{5,6}(q) \=
   \chi_{1,1}^{2,5}(q^2)  \, ,  & \\
 & \!\!\! \chi_{2,1}^{6,7}(q^2) + \chi_{2,6}^{6,7}(q^2) =
   \chi_{1,3}^{6,7}(q) - \chi_{1,4}^{6,7}(q)  , \quad
  \chi_{2,2}^{6,7}(q^2) + \chi_{2,5}^{6,7}(q^2) =
   \chi_{1,1}^{6,7}(q) - \chi_{1,6}^{6,7}(q)  , & \\
 & \chi_{2,3}^{6,7}(q^2) + \chi_{2,4}^{6,7}(q^2) \=
   \chi_{1,2}^{6,7}(q) - \chi_{1,5}^{6,7}(q) \, .  &
\ea

Finally, it may be of some interest to consider relations
between (combinations of) incomplete characters with rescaled $q$.  
For instance, we have
\be{qh}
  \hat{\chi}_{n,m}^{4n,t}(q^2) - \hat{\chi}_{3n,m}^{4n,t}(q^2) \=
    \hat{\chi}_{n,2m}^{2n,t}(q) \, , \quad
 \hat{\chi}_{n,m}^{6n,t}(q^2) - \hat{\chi}_{5n,m}^{6n,t}(q^2) \=
   \hat{\chi}_{n,2m}^{3n,t}(q) \, .
\ee
Identities between the corresponding full characters are then
obtained by multiplication of the r.h.s.~with 
$ (q^{\frac{1}{24}} \{1 \}_1^+ )^{-1} $.

\subsection{Rogers-Ramanujan Type Identities} \

\noindent
Once we have achieved factorization of (combinations of) characters
in the form (\ref{PPP}), we can employ (\ref{notinv}) in order to
re-express the product as a sum of  type distinct from (\ref{char}). 
More precisely, 
combining (\ref{notinv}) with (\ref{defx}) and substituting into
(\ref{PPP}), we obtain
\begin{equation} \label{oi}
 q^{\rm const } \, \frac{\prod_{j=1}^M \{\tilde{x}_i\}^+_{y} }{ 
  \prod_{i=1}^N \{x_j\}^-_{y}    } \= 
 \sum\limits_{\vec{l}}\frac{q^{(l_{1}^{2}+\ldots +l_{M}^{2}-l_{1}
 -\ldots -l_{M})y/2+\vec{B}\cdot \vec{l}}}{(q^{y})_{l_{1}  } \ldots
 (q^{y})_{l_{M+N}}} \,,
\end{equation}
where $\vec{B}=\{ \tilde{x}_{1},\ldots 
 \tilde{x}_{M}, x_{1}, \ldots x_{N} \}$
and $\vec{l}$ has $(M+N)$ components running through non-negative 
integers. The structure of this identity resembles the famous 
Rogers-Ramanujan identities (which are in fact just two ways of 
writing down the characters $ \chi_{1,1}^{2,5}$ and 
$\chi_{1,2}^{2,5}$ -- see appendix A)
\be{rr}
 \frac{1}{ \{1;4\}_{5}^{-} } \= 
 \sum_{l=0}^\infty \frac{q^{l^2}}{(q)_l} \, ,\quad\quad
 \frac{1}{ \{2;3\}_{5}^{-} } \= 
 \sum_{l=0}^\infty \frac{q^{l^2+l}}{(q)_l} \ .
\ee

However, whereas eq.~(\ref{oi}) may be decomposed into a product 
of identities (\ref{notinv}), such simplifications are not possible 
in the  proof of the Rogers-Ramanujan identities (see e.g.~\cite{Har}). 
Thus, in order to obtain more interesting generalizations of the 
Rogers-Ramanujan identities involving our factorized form of 
(combinations of) characters as a product side, we need another 
expression for the sum on the r.h.s.~of (\ref{oi}). For this purpose 
we make use of the results of \cite{KKMM} where it was observed that
some Virasoro characters admit the following form:
\be{kk}
  q^{\rm const } \sum_{\vec{l}} \frac{ q^{\vec{l}^t A \vec{l} + 
  \vec{B}\cdot\vec{l} }  }{ (q)_{l_1}\ldots (q)_{l_n}} \ ,
\ee
where $A$ is a real $n\times n$ symmetric matrix (sometimes 
coinciding with the inverse Cartan matrix of a simply-laced Lie 
algebra), $\vec{B}$ is an $n$-component vector, and the summation 
may be restricted by a condition of the type 
$\vec{\gamma} \cdot \vec{l} = Q\, (\rm mod\, \alpha)$ with some
integer valued $\vec{\gamma}$ and positive $Q$ and $\alpha$.
It turns out that some of the characters of minimal models
admitting the form (\ref{kk}) are either factorizable per se or can 
be combined into the factorizable combinations considered above.
This circumstance allows us to apply the results of section 2 and
derive a set of Rogers-Ramanujan type identities. For instance
\ba 
 \label{rr1}
 q^{ -\frac{1}{40}} \chi_{1,1}^{3,5}(q) &=& 
 \sum_{{l=0}\atop {\rm even}}^\infty 
 \frac{ q^{(l^2+2l)/4} }{(q)_l} \=
 \frac{ \{4\}^+_{10} \{6\}^+_{10} }{ \{2\}^-_{10} \{3\}^-_{10} 
 \{5\}^-_{10} \{7\}^-_{10} \{8\}^-_{10}} \, , \\
 \label{rr2}
 q^{ -\frac{1}{40}} \chi_{1,4}^{3,5}(q) &=& 
 \sum_{{l=1}\atop {\rm odd}}^\infty 
 \frac{ q^{(l^2+2l)/4} }{(q)_l} \= q^{ \frac{3}{4}}
 \frac{ \{1\}^+_{10} \{9\}^+_{10} }{ \{2\}^-_{10} \{3\}^-_{10} 
 \{5\}^-_{10} \{7\}^-_{10} \{8\}^-_{10}} \, .
\ea
Furthermore, we can apply (\ref{3sum}) to combinations of 
the l.h.s.~which yields
\be{rr3}
 q^{ -\frac{1}{40}} \bigl( \chi_{1,1}^{3,5}(q) \pm 
    \chi_{1,4}^{3,5}(q) \bigl) \=
 \sum_{l=0}^\infty \frac{(\pm)^{l}\, q^{(l^2+2l)/4} }{(q)_l} \=
 \frac{ \{3/4\}^\pm_{5/2} \{7/4\}^\pm_{5/2} }{ \{2\}^-_{5} \{3\}^-_{5} 
 \{5/2\}^+_{5/2} } \, .
\ee
We present a set of further Rogers-Ramanujan type identities 
derived in a similar way in appendix E. The product sides of these 
identities are not unique in the sense that one may use the 
techniques discussed in subsection 2.4 and bring them, if possible, 
to the form (\ref{P}), (\ref{PP}) or (\ref{PPP}) (compare (\ref{rr3}) 
and the corresponding formula in appendix A).
It is also worth noticing that, combining these identities further,
we again obtain identities of the Rogers-Ramanujan type. For instance,
multiplying (\ref{rr1}) and (\ref{rr2}), we find
\be{rr4}
 \sum_{l_1,l_2=0}^\infty 
 \frac{ q^{l_1^2+l_2^2+l_1+2l_2} }{ (q)_{2l_1} (q)_{2l_2+1} } \=
 \frac{ \{1\}^+_{5} \{4\}^+_{5} \Bigl( \{5\}^+_{5} \Bigr)^2 }{ 
 \Bigl( \{2\}^-_{5} \{3\}^-_{5} \Bigr)^2 } \, .
\ee

It should be mentioned that there exists a more general type of 
formulae than (\ref{kk}) (involving a $q$-deformed binomial factor) 
\cite{Berk} which covers the whole range of characters in all minimal 
models. Therefore when our factorization technique applies we also 
have Rogers-Ramanujan identities for these more general types.

\subsection{Quasi-particle Representation} \

\noindent
Once a character admits a factorizable form, it is easy to obtain a
quasi-particle spectrum following the prescription of \cite{KKMM,BF,BF1}. 
Let ${\cal{P}}(n,m)$ be the number of partitions of a positive integer 
$n$ into $m$ distinct non-negative integers and ${\cal{Q}}(n,m)$ be the 
number of partitions of a positive integer $n$ into positive integers 
smaller or equal to $m$. The following formulae are well known 
in the number theory (e.g.~\cite{Har}): 
\begin{equation}
 \sum_{n=0}^{\infty } {\cal{P}}(n,m)\,q^{n} \= 
 \frac{q^{m(m-1)/2}}{(q)_{m}} \,,\quad \quad \quad 
 \sum_{n=0}^{\infty }{\cal{Q}}(n,m)\,q^{n}= \frac{1}{(q)_{m}} \,.
\end{equation}
Combining them with (\ref{notinv}) and (\ref{defx}), we obtain 
\begin{eqnarray}
 \{x\}_{y}^{+} &=&\sum_{n,m=0}^{\infty }{\cal{P}}(n,m)\,q^{ny+mx} \=
 \sum_{n,m=0}^{\infty }{\cal{Q}}(n,m)\,q^{(n+m(m-1)/2)y+mx}, \\
 \quad \quad \frac{1}{\{x\}_{y}^{-}} &=&
 \sum_{n,m=0}^{\infty }{\cal{Q}}(n,m)\,q^{ny+mx} \=
 \sum_{n,m=0}^{\infty }{\cal{P}}(n,m)\,q^{(n-m(m-1)/2)y+mx}.
\end{eqnarray}
We assume now the character to be of the form (\ref{oi}),
and  proceed  in the usual way in order to derive the
quasi-particle states. For this  one interprets the characters as a
partition function with 
$\chi (q=e^{-2\pi v/LkT})=\sum_{l=0}^{\infty}P(E_{l})e^{-E_{l}/kT}$, 
$k$ being Boltzmann's constant, $T$ the temperature, $L$ the size of
the quantizing system, $v$ the speed of sound, $E_{l}$ the energy of a
particular level and $P(E_{l})$ its degeneracy. The contribution of a
single particle of type $a$ and momentum $p_{a}^{i_{a}}$ ($i_{a}$ 
being an additional internal quantum number) to the energy is assumed
to be of the form 
$E_{l}=v\sum_{a=1}^{N+M}\sum_{i_{a}=1}^{l_{a}}\left|
p_{a}^{i_{a}}\right|$. 
One has now the option to construct either a purely fermionic 
(in units of $2 \pi /L $)
\begin{equation}
 p_{a}^{i}(\, \vec{l} \,) 
\=B_{a}+\frac{y}{2}\left( 1-\sum\limits_{k=M+1}^{N+M}l_{k}
 \right) +yN_{a}^{i}
\end{equation}
or purely bosonic spectrum (in units of $2 \pi /L $)
\begin{equation}
 p_{a}^{i}(\, \vec{l} \,) 
\=B_{a}+\frac{y}{2}\left( 1-\sum\limits_{k=1}^{M}l_{k}
 \right) +yM_{a}^{i}.
\end{equation}
Here $N_{a}^{i}$ are distinct positive integers and $M_{a}^{i}$ are 
some arbitrary integers. The  fermionic nature of this spectrum is 
here expressed through the fact that the integers $N_a^i$ are all 
distinct, such that we have a Pauli principle. An example for such 
spectra is presented in Table 1. A particular
interesting spectrum arises when we allow bosons and fermions
\begin{equation}
 p_{a}^{i}    \= B_{a}+yN_{a}^{i},\quad 
\;\qquad p_{b}^{i}  =B_{b}+yM_{b}^{i}
\end{equation}
with $a\in \{1,M \}$ and $b\in \{M+1,N+M\}$. Notice now the 
dependence on $ \vec{l}$ has vanished. When $N=M$ this may be
interpreted in a super-symmetric way.

Following the procedure of this subsection, 
the answer to the question \cite{BM}:  ``How many fermionic
representations are there for the characters  of each model 
${\cal M}(s,t)$?'' would be {\it infinite}  for factorizable
characters due to the second relation in (1.14).
One could also change
the approach and start with a given spectrum and search for the
related character \cite{BMP} which shifts the problem to finding 
all possible integrable lattice models. A possible selection mechanism 
is given by using information from the massive models which in the 
conformal limit lead to certain models ${\cal M}(s,t)$. In this spirit 
for instance the choice  $A_1$ and $E_8$ for the algebras of the 
related Cartan matrices in (\ref{kk}) appears quite natural.

\begin{table} 
\begin{center}
\begin{tabular}{|l|l|l|l|}
\hline
k & $ \mu_k $ & $p^{i}=1+2M_{i}$ & $p^{i}(l)=(2-l)+2N_{i}$ \\ \hline
1 & 1 & $\left| 1\right\rangle $ & $\left| 1\right\rangle $ \\ \hline
2 & 1 & $\left| 1,1\right\rangle $ & $\left| 0,2\right\rangle $ \\ \hline
3 & 2 & $\left| 1,1,1\right\rangle ,\left| 3\right\rangle $ & $\left|
-1,1,3\right\rangle ,\left| 3\right\rangle $ \\ \hline
4 & 2 & $\left| 1,1,1,1\right\rangle ,\left| 1,3\right\rangle $ & $\left|
-2,0,2,4\right\rangle ,\left| 0,4\right\rangle $ \\ \hline
5 & 3 & $\left| 1,1,1,1,1\right\rangle ,\left| 1,1,3\right\rangle ,\left|
5\right\rangle $ & $\left| -3,-1,1,3,5\right\rangle ,\left|
-1,1,5\right\rangle ,\left| 5\right\rangle $ \\ \hline
6 & 4 & $
\begin{array}{l}
\left| 1,1,1,1,1\right\rangle ,\left| 3,3\right\rangle , \\ 
\left| 1,5\right\rangle ,\left| 1,1,1,3\right\rangle 
\end{array}
$ & $
\begin{array}{l}
\left| -2,0,2,6\right\rangle ,\left| 0,6\right\rangle , \\ 
\left| -4,-2,0,2,4,6\right\rangle ,\left| 2,4\right\rangle 
\end{array}
$ \\ \hline
7 & 5 & $
\begin{array}{l}
\left| 1,1,1,1,1,1\right\rangle ,\left| 1,3,3\right\rangle , \\ 
\left| 1,1,5\right\rangle ,\left| 1,1,1,1,3\right\rangle ,\left|
7\right\rangle 
\end{array}
$ & $
\begin{array}{l}
\left| -5,-3,-1,1,3,5,7\right\rangle ,\left| -1,1,7\right\rangle , \\ 
\left| -1,3,5\right\rangle ,\left| -3,-1,1,3,7\right\rangle ,\left|
7\right\rangle 
\end{array}
$ \\ \hline
\end{tabular}
\end{center}
\caption{ 
Bosonic and fermionic spectrum for
$\chi _{1,2}^{3,4}(q)=\frac{q^{1/24}}{\{1\}_{2}^{-}}$.
$k$ denotes the level and $ \mu_k $ its degeneracy.}
\end{table}

\subsection{Super-conformal Characters} \

\noindent 
Linear combinations of characters may be found in various contexts as 
for instance when considering super-conformal theories. The two N=1 
unitary minimal super-conformal extension of the Virasoro algebra are 
characterized by an integer $l$ and a label $s=R,NS$, which refers to 
the Ramond or Neveu-Schwarz sector. The Virasoro central charge was 
found \cite{FQS2} to be 
\begin{equation}
c(l)=\frac{3}{2}\left( 1-\frac{8}{l(l+2)}\right) \,,
 \qquad \qquad l=3,4,\ldots \,.
\label{superc}
\end{equation}
The corresponding irreducible representations are highest weight
representations with weights 
\begin{equation}
H_{n,m}^{l,s}=\frac{((l+2)n-ml)^{2}-4}{8l(l+2)}+\frac{1}{16}\delta
_{s,R\quad ,}
\end{equation}
where the labels are restricted as 
$1\leq n\leq l-1$, $1\leq m\leq l+1$ 
together with  $n-m=$
even, odd when $s=NS$, $s=R$, respectively.
Realizing these models as 
$\hat{SU}(2)_{l-2}\otimes \hat{SU}(2)_{2}/\hat{SU}(2)_{l+2}$-cosets 
the corresponding characters  $\Xi _{n,m}^{l,s}(q)$ were constructed 
in \cite{GKO}. One notices from (\ref{c}) and (\ref{superc}) that 
$c(3)=c(4,5)$ and indeed, applying twice the GKO-sumrules one may 
identify super-symmetric characters with linear combinations of some  
non-supersymmetric Virasoro characters
\begin{eqnarray}
\Xi_{1,1}^{3,NS}(q) &=&\chi_{1,1}^{4,5}(q)+\chi_{1,4}^{4,5}(q) \,,
 \qquad  \Xi_{1,1}^{3,\tilde{NS}}(q)=
 \chi_{1,1}^{4,5}(q)-\chi_{1,4}^{4,5}(q) \,, \\
\Xi_{1,3}^{3,NS}(q) &=&\chi_{1,2}^{4,5}(q)+\chi_{1,3}^{4,5}(q) \,,
 \qquad  \Xi_{1,3}^{3,\tilde{NS}}(q)=
 \chi_{1,2}^{4,5}(q)-\chi_{1,3}^{4,5}(q) \,,  \\
\Xi_{2,1}^{3,R}(q) &=&\chi_{2,1}^{4,5}(q) \,, 
 \qquad \qquad \qquad \,\,\,
\Xi_{2,3}^{3,R}(q)=\chi_{2,2}^{4,5}(q) \,.
\end{eqnarray}
Notice that all these characters factorize (see appendix A for 
the explicit formulae). Moreover, they admit the form (\ref{PPP})
(which is due to Rocha-Caridi \cite{Ro}) as well as the form (\ref{P}) 
(see appendix A). It is interesting that the latter does not appear 
to be manifestly super-symmetric. We observe easily the property for 
these expressions under the T-modular transformation (assuming $y$ 
to be an integer, the effect of this transformation is that 
$\{x \}^{\pm}_y \rightarrow  \{x \}^{\pm}_y  $ when $x$ is an integer 
and $\{x \}^{\pm}_y \rightarrow  \{x \}^{\mp}_y $ when $x$ is a 
semi-integer) which relates $\Xi _{n,m}^{l,NS}(q)$ and 
$\Xi _{n,m}^{l,\tilde{NS}}(q)$ and leaves $\Xi _{n,m}^{l,R}(q)$
invariant. Fermionic representations for all characters  
$\Xi _{n,m}^{l,s}(q)$ were found in \cite{Gepn} and we leave it for 
future investigations to settle the question whether they also 
factorize or not. As in the non-supersymmetric case the modular 
properties of these characters \cite{sumod} will certainly turn out 
to be useful.

\subsection{Modular invariant Partition Functions} \

\noindent 
Modular invariant partition functions for minimal models 
are given by (up to an overall coefficient)
\be{zp}
  Z^{s,t}(q,\bar{q}) \= \sum_{n,n',m,m'}  Z_{n,n'}^{m,m'} \, 
 \chi_{n,m}^{s,t}(q) \, \overline{\chi_{n',m'}^{s,t}(q)} \, .
\ee
For the so-called main sequence (in the terminology of \cite{itz}),
or $(A_{s-1},A_{t-1})$ type, we have 
$Z_{n,n'}^{m,m'} = \delta_{n,n'} \delta_{m,m'}$.
Bearing in mind factorizability of all characters in the 
${\cal M}(2,t)$ and ${\cal M}(3,t)$ models, one can rewrite
the corresponding partition functions as a sum of products
of the type (\ref{P}). This allows, in particular, to apply the
technique of subsection 3.3 and obtain quasi-particle 
representations for these  partition functions.

Besides the main sequence some minimal models possess other
modular invariants (complementary sequences) \cite{cap,itz,gep}
of the type (\ref{zp}) with more general $Z_{n,n'}^{m,m'}$. 
In particular, for ${\cal M}(4k,t)$ and ${\cal M}(4k+2,t)$
($(D_{2k+1},A_{t-1})$ and $(D_{2k+2},A_{t-1})$ type, respectively)
the non-diagonal part of $Z_{n,n'}^{m,m'}$ is
$z_{nm} \delta_{n,n'} \delta_{m,t-m'}$. In this case (\ref{zp})
involves not only squares of modules of single characters but also
those of sums of characters of the type (\ref{com}). For $t=3$ all 
of these sums are factorizable and we can represent
the corresponding partition functions as a sum of products
(of the type (\ref{PPP}) in general). Thus, for such 
partition functions we also can obtain quasi-particle 
representations.

\subsection{Partition Functions in boundary CFT} \

\noindent A partition function of a conformal theory on a manifold
with boundaries, say on a cylinder, is expressed as a sum of
characters of a single copy of the Virasoro algebra \cite{Car2}
\be{bz}
 Z_{\alpha,\beta}(q) \= \sum_h N^h_{\alpha \beta} \chi_h(q) \, ,
\ee
where $(\alpha, \beta)$ is a pair of boundary conditions, $\chi_h(q)$
denotes a character of given weight $h$, and $N^h_{\alpha \beta}$
are multiplicities (expressible in terms of (\ref{Sm}) and also 
related to the fusion rules). 

It is interesting that in some cases $Z_{\alpha, \beta}(q)$ is just
a factorizable sum (or several such sums) of type (\ref{com}), so we
can rewrite it in the product form. For instance, for the critical
3-state Potts model (corresponding to ${\cal M}(5,6)$) there are
three microscopic states A, B and C, and for some of possible
partition functions we find
\ba
  Z_{A,F}(q) &=& \chi^{5,6}_{1,2}(q) + \chi^{5,6}_{4,2}(q) \=
   q^{\frac{11}{120}} \, \frac{ 1 }{ \{1\}_{5/2}^{-} \{3/2\}_{5/2}^{-}
   } \, ,  \label{zp1} \\  
 Z_{BC,F}(q) &=& \chi^{5,6}_{2,2}(q) + \chi^{5,6}_{3,2}(q) \=
   q^{-\frac{1}{120}} \, \frac{ 1 }{ \{1/2\}_{5/2}^{-} \{2\}_{5/2}^{-}
  } \, ,  \label{zp2}  
\ea
where F stands for the free boundary condition. As we mentioned in 
subsection 2.3, such an expression may be interpreted as a character 
of a module generated by bosonic operators (in fact, (\ref{xxx}) 
shows that (\ref{zp1}) and (\ref{zp2}) coincide with the characters of 
${\cal M}(2,5)$ of an argument $q^{1/2}$). Also, this form of a partition 
function allows for a direct extraction of a quasi-particle spectrum 
which, (in the spirit of subsection 3.3) in particular, can be used 
to study connections between theories with distinct boundary conditions.

\section*{Conclusion} \

\noindent
We have shown how to obtain the factorized form of a single 
Virasoro character on the base of the Gau\ss -Jacobi and Watson
identities by exploiting the quasi-classical asymptotics of the 
usual sum representation. We have applied this method also to the 
factorization of a linear combination of two Virasoro characters 
and found the explicit formulae (2.23), (2.26) and (2.27). We
presented a rigorous proof that besides the obtained expressions no
other differences of two Virasoro characters of the type (\ref{com}) 
are factorizable in the form (\ref{P}). It is a remarkable fact, 
which certainly needs some deeper understanding, that just like for 
the single characters none of the Macdonald identities, other than the 
ones corresponding to the $A_1^{(1)}$ and $A_2^{(2)}$ algebras need to 
be invoked. We employed the obtained factorized versions of the 
characters in order to derive a  set of new identities, e.g. 
(\ref{id1})-(\ref{id4}), in a very economical way. Some particular 
cases of these identities coincide with formulae derived originally 
in \cite{Tao}, however now the proof has simplified considerably. 
As was already pointed out in \cite{Tao}, these identities belong to 
a class which is closely related, but not derivable, from a repeated 
use of the GKO-sumrules \cite{GKO}. It is therefore suggestive to 
assume that the new identities are related to some higher sumrules. 
A systematic classification of identities obtainable {}from factorised 
combinations of Virasoro characters will be presented elsewhere. 
It is also conceivable, that the presented method will be applicable 
to non-minimal models like parafermionic models, 
i.e.~$\hat{SU}(2)_k/\hat{U}(1)_k$-coset, or general N=1,2,4 
super-symmetric models. Concerning the quasi-particle representation 
of the Virasoro characters with their relation to lattice models, 
the factorized versions constitute a suitable starting point for a 
more detailed analysis, as for instance in \cite{BF}.

\subsection*{Acknowledgments} \

We would like to thank W.~Eholzer, K.P.~Kokhas, C.~Korff, B.M.~McCoy,
V.~Schomerus,  M.A.~Semenov-Tian-Shanski and O.~Verhoeven
for useful discussions and remarks. 
A.B.~is grateful to the members of the Institute f\"ur
Theoretische Physik, FU-Berlin for their hospitality and to the 
Volkswagen Stiftung for partial financial support. 
A.F.~is grateful to the members of the Steklov Mathematical Institute
(St. Petersburg) for their 
hospitality and to the Deutsche Forschungsgemeinschaft
(Sfb288) for partial financial support.

\setcounter{section}{0}
\renewcommand{\thesection}{}
\renewcommand{\theequation}{\Alph{section}.\arabic{equation}}

\section{Appendix A}

\noindent
Here we will present some examples of the inverse product 
representation for characters and linear combinations of
characters in some unitary and non-unitary models. For
shortness we omit the argument $q$ on l.h.s.~and use
the notation
$\{x_1;\ldots ;x_n\}_y^\pm:=\{x_1\}_y^\pm\ldots \{x_n\}_y^\pm$.
\begin{eqnarray*} 
&& \chi_{1,1}^{3,4} \pm \chi_{1,3}^{3,4}  =
 q^{-\frac{1}{48}} \frac{1}{ \{1/2\}_{1}^{\mp} \{1\}_{1}^{+}} \, ,
 \quad\quad  \chi_{1,2}^{3,4}  =
 q^{\frac{1}{24}} \frac{1}{ \{1\}_{2}^{-} } \, , \\
&& \chi_{2,1}^{4,5} = q^{\frac{49}{120}}
 \frac{1}{ \{1;4\}_{5}^{-} \{3;5;7\}_{10}^{-}  } \, ,
 \quad\quad    \chi_{2,2}^{4,5} = q^{ \frac{1}{120}}
 \frac{1}{ \{2;3\}_{5}^{-} \{1;5;9\}_{10}^{-}  } \, , \\
&& \chi_{1,1}^{4,5} \pm \chi_{1,4}^{4,5} =
 q^{-\frac{7}{240}} \frac{1}{ \{3/2;5/2;7/2\}_{5}^{\mp}
 \{5\}_{5}^{+} \{2;8\}_{10}^{-} } \, ,\\
&& \chi_{1,2}^{4,5} \pm \chi_{1,3}^{4,5} =
 q^{ \frac{17}{240}} \frac{1}{ \{1/2;5/2;9/2\}_{5}^{\mp}
 \{5\}_{5}^{+} \{4;6\}_{10}^{-} } \, ,\\
&& \chi_{1,1}^{5,6} - \chi_{1,5}^{5,6} =
 q^{-\frac{1}{30}} \frac{1}{ \{2;8\}_{10}^{-} } \, ,
 \quad\quad
 \chi_{2,1}^{5,6} - \chi_{2,5}^{5,6} =
 q^{ \frac{11}{30}} \frac{1}{ \{4;6\}_{10}^{-} } \, ,\\
&& \chi_{1,2}^{5,6} \pm \chi_{1,4}^{5,6} =
  \frac{ q^{\frac{11}{120}} }{ \{1;4\}_{5}^{-} 
  \{3/2;7/2\}_{5}^{\mp} } \, , \quad\quad
 \chi_{2,2}^{5,6} \pm \chi_{2,4}^{5,6} =
  \frac{ q^{-\frac{1}{120}} }{ \{2;3\}_{5}^{-} 
  \{1/2;9/2\}_{5}^{\mp} } \, ,  \\
&& \chi_{1,1}^{6,7} - \chi_{1,6}^{6,7} =
  \frac{ q^{-\frac{1}{28}} }{ \{3;4\}_{7}^{-} 
  \{2;12\}_{14}^{-} } \, , \quad\quad
 \chi_{1,2}^{6,7} - \chi_{1,5}^{6,7} =
  \frac{ q^{\frac{3}{28}} }{ \{1;6\}_{7}^{-} 
  \{4;10\}_{14}^{-} } \, , \\
&& \chi_{1,3}^{6,7} - \chi_{1,4}^{6,7} =
  \frac{ q^{ \frac{19}{28}} }{ \{2;5\}_{7}^{-} 
  \{6;8\}_{14}^{-} } \, ,  \\
&& \chi_{2,1}^{6,7} \pm \chi_{2,6}^{6,7} =
  \frac{ q^{\frac{19}{56}} }{ \{1;3;4;6\}_{7}^{-} 
  \{5/2;9/2\}_{7}^{\mp} }  \, ,  \quad\ 
 \chi_{2,2}^{6,7} \pm \chi_{2,4}^{6,7} =
  \frac{ q^{-\frac{1}{56}} }{ \{1;2;5;6\}_{7}^{-} 
  \{3/2;11/2\}_{7}^{\mp} }  \, ,  \\
 && \chi_{2,3}^{6,7} \pm \chi_{2,4}^{6,7} =
   q^{\frac{3}{56}} \frac{1}{ \{2;3;4;5\}_{7}^{-} 
  \{1/2;13/2\}_{7}^{\mp} }  \, , \\
 && \chi_{1,1}^{2,5} = q^{\frac{11}{60}}
 \frac{1}{ \{2;3\}_{5}^{-} } \, , \quad\quad  
 \chi_{1,2}^{2,5} = q^{-\frac{1}{60} }
 \frac{1}{ \{1;4\}_{5}^{-} } \, ,\\
 && \chi_{1,1}^{3,5} \pm  \chi_{1,4}^{3,5} =
 q^{\frac{1}{40}}
 \frac{1}{ \{2;8\}_{10}^{-} \{3/4;7/4\}_{\frac 52}^{\mp}
 \{3/2;5/2;7/2\}_{5}^{+}} \, , \\
 && \chi_{1,2}^{3,5} \pm  \chi_{1,3}^{3,5} =
 q^{-\frac{1}{40}} 
 \frac{1}{ \{4;6\}_{10}^{-} \{1/4;9/4\}_{\frac 52}^{\mp}
 \{1/2;5/2;9/2\}_{5}^{+}} \, .
\end{eqnarray*} 

\section{Appendix B}

\noindent
In this appendix we present a sample proof for the identities of 
the type (\ref{4p})-(\ref{4m}) and (\ref{3sum})-(\ref{6sum}), 
that is for the factorization of the sum 
or difference of two Virasoro characters related to minimal models. 
The proof is based on a systematic exploitation of 
the Gau\ss-Jacobi and Watson identities (\ref{GJ1})-(\ref{W2}).
We have to compare the l.h.s.~of these expressions with the sum or 
difference of characters given by (\ref{char}),
\begin{eqnarray}
 \chi_{n,m}^{s,t} (q) \pm \chi _{n,t-m}^{s,t} (q) 
 &=&\frac{q^{h_{n,m}^{s,t}-\frac{c(s,t)}{24}}}
 {(q)_{\infty }}\sum_{k=-\infty }^{\infty }q^{stk^{2}}\left(
 q^{k(nt-ms)}-q^{k(nt+ms)+nm}\right.   \nonumber \\
 &&\left. \pm \, q^{k(nt+ms-st)+ \D h_{n,m}^{s,t} }
 \mp q^{k(nt-ms+st)+ n(t-m)+ \D h_{n,m}^{s,t}} \right)  .  
 \label{minsum}
\end{eqnarray}
Here the quantity $\D h_{n,m}^{s,t}$
is defined by (\ref{dhp}) and we assume $n < s/2$, $m < t/2$, 
so that $\D h_{n,m}^{s,t} > 0$.
We  outline the proof for the identity (\ref{6sum}). 
All other proofs work along the same lines. 

Recall that (\ref{6sum}) has been conjectured to be a particular
case of (\ref{W1}) for $a = \D h_{n,m}^{s,t}$ and $b=nm$ 
provided that the condition $s=6n$ holds. Notice that substitution 
of the latter relation into (\ref{dhp}) yields $a=nt -2nm$.
In order to produce the right number of terms for a possible
comparison with (\ref{minsum}), we have to split the sum in the 
l.h.s.~of (\ref{W1}) into two new sums -- over even and odd $k$. 
Then the l.h.s.~of (\ref{W1}) acquires the form
\[
 \sum_{k=-\infty }^{\infty} q^{k^{2}(6a+12b)} \left(q^{k(a-4b)}
 + q^{k(7a+8b)+2a+b} - q^{k(a+8b)+b}-q^{k(7a+20b)+2a+8b}\right) \, ,
\]
which, upon substitution of the explicit values for $a$ and $b$
and the relation $s=6n$, becomes
\[
 \sum_{k=-\infty }^{\infty} q^{st k^{2}} \left(q^{k(nt-ms)}
 + q^{k(nt-ms+st)+ 2nt-3nm} - q^{k(nt+ms)+nm}-
 q^{k(nt+ms+st)+2nt+4nm}\right) \, .
\]
We see that the first, second and third terms here exactly match 
the first, fourth and second terms on the r.h.s. in (\ref{minsum}), 
respectively. 
Making the shift $k \rightarrow k-1$ in the last term, we achieve 
that it coincides with the third term in (\ref{minsum}). 
This completes the proof.

\section{Appendix C}

\noindent
Here we will prove the following statement: the factorization of the 
difference of two minimal Virasoro characters in the form
\begin{equation}
 \chi_{n,m}^{s,t}(q) - \chi_{s-n,m}^{s,t}(q) =
 \frac{q^{h_{n,m}^{s,t}-\frac{c(s,t)}{24}}} {(q)_{\infty }}
 \left( \hat{\chi}_{n,m}^{s,t}(q) -q^{\Delta h}
 \hat{\chi}_{s-n,m}^{s,t}(q) \right) = 
 \frac{ q^{h_{n,m}^{s,t}-\frac{c(s,t)}{24}} }{
 \prod_i^N \{x_i\}_{b}^{-}} \, ,  \label{theo}
\end{equation}
where $0 < x_1 < \ldots < x_N \leq b$,
is up to the symmetries (\ref{hsym}) only possible for $s=3n,4n,6n$.
Here $\D h$ stands for $\D h_{n,m}^{s,t}$ defined in (\ref{dhp}),
and we assume $n < s/2$, $m < t/2$, so that $\D h_{n,m}^{s,t} > 0$.

Our argumentation goes along the lines of the
proof for the factorization of single characters given 
in \cite{chr}. Surprisingly it is enough to investigate the first 
five terms in the sum, which for the incomplete character may be 
identified uniquely
\begin{equation}
\hat{\chi}_{n,m}^{s,t} (q) = 1 - q^{nm} - q^{(s-n)(t-m)} + 
 q^{ts+sm-tn} + q^{ts+tn-sm} +  \ldots \,.
\end{equation}
For the difference of the two characters they read
\be{cc}
 \hat{\chi}_{n,m}^{s,t}(q) - q^{\Delta h}\hat{\chi}_{s-n,m}^{s,t}(q)
  = 1-q^{nm}-q^{\Delta h}+q^{\Delta h+m(s-n)}+q^{\Delta h+n(t-m)}
  + \ldots \,.
\ee
For definiteness we choose $sm<nt$ (so that 
$\Delta h+m(s-n) < \Delta h+n(t-m)$), since the other case may be 
obtained from the symmetry properties.
The negative terms in (\ref{cc}) allow us to write down the first 
two factors in the product
\begin{equation}
 \hat{\chi}_{n,m}^{s,t}(q) - q^{\Delta h} 
 \hat{\chi}_{s-n,m}^{s,t}(q) =
 (1-q^{nm})(1-q^{\Delta h}) \ldots \,,  \label{fac}
\end{equation}
which means that after expanding we will generate a term 
$q^{nm+\Delta h}$. Since $nm+\Delta h < \Delta h+m(s-n)$,
we have to include a factor $(1-q^{nm+\Delta h})$ on the r.~h.~s.~of 
(\ref{fac}) in order to cancel this term. Expanding once more we will 
generate new terms, which in turn have to be cancelled by additional 
factors on the right 
hand side of (\ref{fac}) until we obtain the matching condition 
$\alpha nm+\beta \Delta h=\Delta h+m(s-n)$   
with positive integers $\alpha $ and $\beta$.
At first sight it seems a formidable task to bring some systematics 
into this analysis. However, it was observed in \cite{chr} that 
this procedure will terminate when $\alpha+\beta =5$. Actually also 
one case from level $6$ might be possible.

Performing this analysis up to that level, one obtains
\begin{eqnarray*}
 && 1-q^{nm}-q^{\Delta h}+q^{\alpha nm+\beta \Delta h} +\ldots \= \\
 && \quad\quad (1-q^{nm})(1-q^{\Delta h}) (1-q^{nm+\Delta h})
 (1-q^{2nm+\Delta h}) \\
 && \quad\quad  \times (1-q^{nm+2\Delta h})(1-q^{3nm+\Delta h})
 (1-q^{nm+3\Delta h}) (1-q^{2nm+2\Delta h}) \\  
 && \quad\quad \times (1-q^{4nm+\Delta h})(1-q^{nm+4\Delta h}) 
 (1-q^{3nm+2\Delta h})^{2}(1-q^{2nm+3\Delta h})^{2} \ldots \, ,
\end{eqnarray*}
where in this expression $\alpha +\beta >5.$ It is the occurrence of 
the quadratic terms $(1-q^{3nm+2\Delta h})^{2}$ and 
$(1-q^{2nm+3\Delta h})^{2}$ which allows us to stop at this point, 
since they may never be cancelled against factors within 
$(q)_{\infty }$ and we can therefore restrict the investigation to 
the cases $2\leq \alpha +\beta \leq 5$. Commencing with the
case $\alpha +\beta =5$ we obtain two matching conditions, that is for
the two smallest powers of the positive terms
\[
 \left. \ar{l} \frac{st}{4}+\frac{ms}{2}-\frac{nt}{2}
  =3nm+2\Delta h \\ [1mm]
  \frac{st}{4}-\frac{ms}{2}+\frac{nt}{2}= 2nm+3\Delta h \er
 \right\} \quad  \Rightarrow s=6n-2m \frac{s}{t} \, .
\]

Since $n$ is positive, $m$ is strictly smaller than $t$, \, 
$\langle s,t \rangle =1$ and $t=2m$ produces zero on the left 
hand side of (\ref{theo}), the case $\alpha +\beta =5$ will never 
produce any solution. We may also encounter the situation 
\[
 \left. \ar{l} \frac{st}{4}+\frac{ms}{2}-\frac{nt}{2}
  =4nm+\Delta h \\ [1mm]
  \frac{st}{4}-\frac{ms}{2}+\frac{nt}{2}= 2nm+2\Delta h \er
 \right\} \quad  \Rightarrow s=5n \;\; {\rm and } \;\; t = 6m \, .
\]
In the remaining possibilities we only obtain 
one matching condition, that is for $p$. 
The case $\alpha =\beta =2$ leads to the condition 
$2nm+2\Delta h=sm-nm+\Delta h$ which amounts to
\[
  m=t\frac{(s-2n)}{(6s-16n)} \, .
\]
However, substitution of this relation into the condition
$sm<nt$ leads to $s(s-2n)<n(6s-16n)$, or equivalently,
$(s-4n)^2<0$, which is impossible.

The other cases yield
\begin{eqnarray*}
nm+\Delta h=sm-nm+\Delta h\quad  &\Rightarrow &\quad s=2n \,, \\
nm+2\Delta h=sm-nm+\Delta h\quad  &\Rightarrow &\quad s=2n
\ \ {\rm or }\ \ t=6m \,,\\
2nm+\Delta h=sm-nm+\Delta h\quad  &\Rightarrow &\quad s=3n \,,\\
nm+3\Delta h=sm-nm+\Delta h\quad  &\Rightarrow &\quad s=2n
\ \ {\rm or } \ \ t= 4m \,,\\
3nm+\Delta h=sm-nm+\Delta h\quad  &\Rightarrow &\quad s=4n \,,\\
5nm+\Delta h=sm-nm+\Delta h\quad  &\Rightarrow &\quad s=6n \,.
\end{eqnarray*}
We observe that we recover the cases we claimed to factorize in 
the form (\ref{fac}), which concludes the proof.

\section{Appendix D}

\noindent
We will now provide a sample proof for the identities 
(\ref{id1})-(\ref{id4}). For $n=m=1$ some very involved proof which 
employs identities of theta functions may be found in \cite{Tao}. 
With the help of the product representations (\ref{4p})-(\ref{6sum}) 
such identities may be derived without any effort. We demonstrate 
this just for equations (\ref{id3}) with the upper sign, the  
remaining equations may be derived in a similar way. First of all 
we notice that 
\begin{equation}
\label{D1}
 h_{n,m}^{3n,2m} - \frac{c({3n,2m})}{24}+h_{n,m}^{4n,5m} -
 \frac{c({4n,5m})}{24} = h_{2n,2m}^{6n,5m} -
 \frac{c({6n,5m})}{24}+h_{n,m}^{3n,4m}-\frac{c({3n,4m})}{24} \, .
\end{equation}
After cancelling $(q)_{\infty }^{2}$ on both sides of (\ref{id3}) for 
the upper sign we obtain for the left-hand side upon using (\ref{pr1}) 
(we omit here the labels  $nm$ in order to avoid lengthy formulae and 
imagine just for now that $\{x\}_{y}^{-}$ should always be understood 
as $\{xnm\}_{ynm}^{-}$) 
\begin{eqnarray*}
 && \hat{\chi}_{n,m}^{3n,2m}\left( \hat{\chi}_{n,m}^{4n,5m}+
 \hat{\chi}_{n,4m}^{4n,5m}\right)  \\
 &=& \left( \{1\}_{1}^{-} \right) \left(\{1\}_{5}^{-}\{4\}_{5}^{-}
 \{5\}_{5}^{-} \left\{ \frac{3}{2}\right\}_{5}^{+}\left  
 \{\frac{5}{2}\right\}_{5}^{+}\left\{ \frac{7}{2}\right\}_{5}^{+} 
\right) \\
 &=&\{1\}_{4}^{-}\{3\}_{4}^{-}\{2\}_{2}^{-}\{1\}_{10}^{-}
 \{6\}_{10}^{-}\{4\}_{10}^{-}\{9\}_{10}^{-}\{5\}_{5}^{-}
 \left\{ \frac{3}{2}\right\}_{5}^{+}\left\{ \frac{5}{2}
 \right\}_{5}^{+}\left\{ \frac{7}{2}\right\}_{5}^{+} \\
 &=& \left( \{1\}_{4}^{-}\{3\}_{4}^{-}\{2\}_{2}^{-}\left\{ \frac{1}{2}
 \right\}_{2}^{+}\left\{ \frac{3}{2}\right\}_{2}^{+} \right) \left(
 \{4\}_{10}^{-}\{6\}_{10}^{-}\{5\}_{5}^{-}\left\{ \frac{1}{2}
 \right\}_{5}^{-}\left\{ \frac{9}{2}\right\} _{5}^{-} \right) \\
 &=& \left( \hat{\chi}_{n,m}^{3n,4m}+\hat{\chi}_{n,3m}^{3n,4m}\right)
\left( \hat{\chi}_{2n,2m}^{6n,5m}-\hat{\chi}_{2n,3m}^{6n,5m}\right) \, . 
\end{eqnarray*}
Here we have used several times the identities (\ref{pr1}).

\section{Appendix E}

\noindent
We complement the list started in subsection 3.2 of the 
Rogers-Ramanujan type identities obtained by combining our product 
formulae for (combinations of) characters with the results of 
\cite{KKMM}. We adopt the notations explained in appendix A.
\ba
 \label{RR1}  q^{1/48} \chi_{1,1}^{3,4} &=& 
   \sum_{{l=0}\atop{\rm even}}^\infty \frac{ q^{l^2/2} }{(q)_l} =
   \frac{1}{  \{2;14\}^-_{16} \{3;4;5\}^-_{8} } \, , \\
 \label{RR2}  q^{1/48} \chi_{1,3}^{3,4} &=&
 \sum_{{l=1}\atop{\rm odd}}^\infty \frac{ q^{l^2/2} }{(q)_l} =
 q^{1/2} \frac{1}{  \{6;10\}^-_{16} \{1;4;7\}^-_{8} } \, , \\
 \label{RR3} q^{1/48} \Bigl( \chi_{1,1}^{3,4} \pm 
  \chi_{1,3}^{3,4} \Bigr) &=& \sum_{l=0}^\infty  
 \frac{ (\pm)^{l}\, q^{l^2/2} }{(q)_l} = \{1/2\}^\pm_{1} \, , \\
 \label{RR4} q^{-1/24} \chi_{1,2}^{3,4} &=&
 \sum_{{l=0}\atop{\rm even}}^\infty \frac{ q^{(l^2-l)/2} }{(q)_l} =
 \sum_{{l=1}\atop{\rm odd}}^\infty \frac{ q^{(l^2-l)/2} }{(q)_l} =
  \frac{1}{  \{1\}^-_{2} } \, , \\
 \label{RR5} q^{1/30} \chi_{1,3}^{5,6} &=& 
 \sum_{{l_1,l_2=0}\atop{l_1+2l_2 = \pm 1 ({\rm mod}\, 3)}}^\infty 
  \frac{ q^{2(l_1^2 + l_1^2 + l_1 l_2)/3} }{(q)_{l_1} (q)_{l_2}} =
 \frac{ q^{2/3}}{  \{1;2\}^-_{3} \{6;9\}^-_{15} } \, , \\
 \label{RR6} q^{ \frac{1}{40}} \chi_{1,2}^{3,5} &=& 
 \sum_{{l=0}\atop {\rm even}}^\infty 
 \frac{ q^{l^2/4} }{(q)_l} \=
 \frac{ \{3;7\}^+_{10} }{ \{1;4;5;6;9\}^-_{10} } \, , \\
 \label{RR7}
 q^{ \frac{1}{40}} \chi_{1,3}^{3,5} &=& 
 \sum_{{l=1}\atop {\rm odd}}^\infty 
 \frac{ q^{l^2/4} }{(q)_l} \=  q^{1/4}
 \frac{ \{2;8\}^+_{10} }{ \{1;4;5;6;9\}^-_{10} } \, , \\
 \label{RR8} q^{ \frac{1}{40}} \Bigl( \chi_{1,2}^{3,5} \pm 
    \chi_{1,3}^{3,5} \Bigl) &=&
 \sum_{l=0}^\infty \frac{(\pm)^{l}\, q^{l^2/4} }{(q)_l} \=
 \frac{ \{1/4\}^\pm_{5/2} \{9/4\}^\pm_{5/2} }{ \{1;4\}^-_{5} 
  \{5/2\}^+_{5/2} } \, .
\ea
More identities will be given elsewhere \cite{AB}.

\newcommand{\sbibitem}[1]{ \vspace*{-1.5ex} \bibitem{#1} }

\end{document}